\documentclass[aps,prd,nofootinbib,twocolumn,showpacs,superscriptaddress]{revtex4-1}

\pdfoutput=1
\usepackage {natbib}
\usepackage {graphicx}
\usepackage {hyperref}
\usepackage {color}
\usepackage{textcase}
\usepackage{amsmath, amsthm, amsfonts}
\usepackage{enumerate}
\usepackage{soul}
\usepackage[normalem]{ulem}

\def\mbf#1{\ensuremath{\mathchoice{\mbox{\boldmath$\displaystyle#1$}}
{\mbox{\boldmath$\textstyle#1$}}
{\mbox{\boldmath$\scriptstyle#1$}}
{\mbox{\boldmath$\scriptscriptstyle#1$}}}}

\renewcommand{\sec}{\mathrm{s}}
\newcommand{\hours}{\mathrm{h}}
\newcommand{\Hz}{\mathrm{Hz}}
\newcommand{\mHz}{\mathrm{mHz}}

\newcommand{\Ord}[1]{\mathcal{O}\left(#1\right)}

\newcommand{\prob}[2]{P\left(#1|#2\right)}
\newcommand{\probI}[1]{P\left(#1\right)}
\newcommand{\scalar}[2]{\langle#1|#2\rangle}

\newcommand{\detVec}[1]{\mbf{#1}}

\newcommand{\prior}[1]{\MakeLowercase{#1}}
\newcommand{\expect}[1]{E\left[ #1 \right]}

\newcommand{\avgX}[1]{\left\langle #1 \right\rangle}	
\newcommand{\avgSeg}[1]{\overline{#1}}			

\newcommand{\OSNpFA}[1]{{}^{(#1)}\OSN}
\newcommand{\utOSNpFA}[1]{{}^{(#1)}\utOSN}
\newcommand{\OSNscpFA}[1]{{}^{(#1)}\OSNsc}

\newcommand{\sig}{\mathrm{s}}
\newcommand{\Dop}{\lambda}
\newcommand{\DopSpace}{\mathbb{P}}
\newcommand{\Freq}{f}
\newcommand{\fdot}{{\dot{\Freq}}}

\newcommand{\Amp}{\mathcal{A}}
\newcommand{\cosi}{\cos\iota}
\newcommand{\phio}{\phi_0}

\newcommand{\rhomax}{{\widehat{\rho}_{\mathrm{max}}}}
\newcommand{\cF}{c_*}

\newcommand{\M}{\mathcal{M}}

\newcommand{\Sn}{S_{\mathrm{n}}}	
\newcommand{\SnX}{\Sn^{X}}		
\newcommand{\SnXal}{\Sn^{X\alpha}}		

\newcommand{\dVx}{\detVec{x}}
\newcommand{\dVy}{\detVec{y}}
\newcommand{\dVn}{\detVec{n}}
\newcommand{\dVh}{\detVec{h}}

\newcommand{\Hyp}{\mathcal{H}}
\newcommand{\Gauss}{\mathrm{\MakeUppercase{G}}}
\newcommand{\Signal}{{\mathrm{\MakeUppercase{S}}}}
\newcommand{\Line}{{\mathrm{\MakeUppercase{L}}}}
\newcommand{\Noise}{{\Gauss\Line}}
\newcommand{\HypS}{\Hyp_\Signal}
\newcommand{\HypN}{\Hyp_\Noise}
\newcommand{\HypL}{\Hyp_\Line}
\newcommand{\HypG}{\Hyp_\Gauss}

\providecommand{\sc}[1]{\widehat{#1}}
\renewcommand{\sc}[1]{\widehat{#1}}
\newcommand{\HypSsc}{\sc{\Hyp}_{\Signal}}
\newcommand{\HypLsc}{\sc{\Hyp}_{\Line}}
\newcommand{\HypNsc}{\sc{\Hyp}_{\Noise}}
\newcommand{\HypGsc}{\sc{\Hyp}_{\Gauss}}

\newcommand{\OSN}{O_{\Signal\Noise}}	
\newcommand{\OSG}{O_{\Signal\Gauss}}	
\newcommand{\OSL}{O_{\Signal\Line}}
\newcommand{\utOSL}{\OSL^{(0)}}
\newcommand{\utOSN}{\OSN^{(0)}}

\newcommand{\OLG}{O_{\Line\Gauss}}
\newcommand{\Fveto}{\F^{\mathrm{+veto}}}
\newcommand{\scFveto}{\scF^{\mathrm{+veto}}}

\newcommand{\Bayes}{B}
\newcommand{\BSG}{\Bayes_{\Signal\Gauss}}

\newcommand{\oSG}{\prior{\OSG}}
\newcommand{\oSL}{\prior{\OSL}}
\newcommand{\oLG}{\prior{\OLG}}
\newcommand{\oSN}{\prior{\OSN}}

\newcommand{\OSNsc}{\sc{O}_{{\Signal\Noise}}}	
\newcommand{\OSGsc}{\sc{O}_{{\Signal\Gauss}}}	
\newcommand{\OSLsc}{\sc{O}_{{\Signal\Line}}}
\newcommand{\utOSLsc}{\OSLsc^{(0)}}

\newcommand{\OLGsc}{\sc{O}_{{\Line\Gauss}}}

\newcommand{\oSLsc}{\prior{\OSLsc}}
\newcommand{\oLGsc}{\prior{\OLGsc}}
\newcommand{\oSGsc}{\prior{\OSGsc}}
\newcommand{\oSNsc}{\prior{\OSNsc}}

\newcommand{\rsc}{\sc{r}}

\newcommand{\lineprob}{p_\Line}
\newcommand{\linefrac}{f_\Line}
\newcommand{\lineprobsc}{\sc{p}_{{\Line}}}

\newcommand{\F}{\mathcal{F}}		

\newcommand{\Fmaxp}{\F'_{\mathrm{max}}}
\newcommand{\Fmaxpp}{\F''_{\mathrm{max}}}

\newcommand{\Fth}{\F_*}
\newcommand{\Ftho}{\Fth^{(0)}}
\newcommand{\FX}{\F^X}
\newcommand{\rX}{r^X}

\newcommand{\scF}{\sc{\F}}
\newcommand{\scFX}{\scF^X}

\newcommand{\scFmaxG}{\scF''_{\mathrm{max}}}
\newcommand{\scFth}{\scF_*}
\newcommand{\scFtho}{\scF_*^{(0)}}
\newcommand{\scrX}{\sc{r}^X}

\newcommand{\avF}{\avgSeg{\F}}

\newcommand{\pDet}{p_{\mathrm{det}}}	
\newcommand{\pFA}{p_{\mathrm{FA}}}
\newcommand{\pFAtho}{p_{\mathrm{FA}*}^{(0)}}
\newcommand{\pFAl}{p_{\mathrm{FA},\Psft}}

\newcommand{\AND}{\;\mathrm{and}\;}
\newcommand{\OR}{\;\mathrm{or}\;}

\newcommand{\Ndet}{{N_{\mathrm{det}}}}
\newcommand{\Nseg}{{N_{\mathrm{seg}}}}

\newcommand{\eto}{\ensuremath{\mathrm{e}^}}	

\newcommand{\segk}{{k}}	

\newcommand{\sft}{{\mathrm{SFT}}}
\newcommand{\Tsft}{T_\sft}

\newcommand{\Psft}{\mathcal{P}} 
\newcommand{\Psftthr}{\Psft_{\mathrm{thr}}} 
\newcommand{\Nsft}{N_\sft}
\newcommand{\Nbins}{N_{\mathrm{bins}}}
\newcommand{\Ncrossed}{N_{\Psft > \Psftthr}}

\newcommand{\LHO}{\textrm{H1}}
\newcommand{\LLO}{\textrm{L1}}

\newcommand{\snr}{\rho} 
\newcommand{\snrS}{\snr_{\Signal}}
\newcommand{\snrL}{\snr_{\Line}}
\newcommand{\tstart}{t_{\mathrm{start}}}

\newcommand{\fsft}{f_\sft} 
\newcommand{\finj}{f_{\mathrm{inj}}}

\newcommand{\asc}{$\sc{\textrm{a}}$}
\newcommand{\bsc}{$\sc{\textrm{b}}$}
\renewcommand{\csc}{$\sc{\textrm{c}}$}
\newcommand{\dsc}{$\sc{\textrm{d}}$}

\newcommand{\acoh}{$\widetilde{\textrm{a}}$}
\newcommand{\bcoh}{$\widetilde{\textrm{b}}$}
\newcommand{\ccoh}{$\widetilde{\textrm{c}}$}
\newcommand{\dcoh}{$\widetilde{\textrm{d}}$}

\newcommand{\maxnoise}[1]{\max#1_{\mathrm{noise}}}

\newcommand{\numexamplebands}{four}

\newcommand{\FvetoName}{$\Fveto$-statistic}
\newcommand{\scFvetoName}{$\scFveto$-statistic}

\newcommand{\dcc}{LIGO-P1300167}
\newcommand{\aei}{AEI-2013-260}

\begin{document}

\title{Search for continuous gravitational waves: Improving robustness versus instrumental artifacts}

\author{David~Keitel}
\email{David.Keitel@aei.mpg.de}
\author{Reinhard~Prix}
\email{Reinhard.Prix@aei.mpg.de}
\affiliation{Albert-Einstein-Institut, Callinstrasse 38, 30167 Hannover, Germany}
\author{Maria~Alessandra~Papa}
\affiliation{Albert-Einstein-Institut, Callinstrasse 38, 30167 Hannover, Germany}
\affiliation{Albert-Einstein-Institut, Am M\"uhlenberg 1, 14476 Golm}
\affiliation{University of Wisconsin-Milwaukee, Milwaukee, Wisconsin 53201, USA}
\author{Paola~Leaci}
\affiliation{Albert-Einstein-Institut, Callinstrasse 38, 30167 Hannover, Germany}
\affiliation{Albert-Einstein-Institut, Am M\"uhlenberg 1, 14476 Golm}
\author{Maham~Siddiqi}
\affiliation{Albert-Einstein-Institut, Callinstrasse 38, 30167 Hannover, Germany}
\affiliation{Harvard-Smithsonian Center for Astrophysics, 60 Garden Street, Cambridge, Massachusetts 02138,
USA}

\date{
published as \href{http://dx.doi.org/10.1103/PhysRevD.89.064023}{Phys. Rev. D 89, 064023}, 10 March 2014. \\
..this version dated 18 February 2014. \\
..LIGO document number: \dcc{}
}

\pacs{
04.30.Tv, 
04.80.Nn, 
95.55.Ym, 
97.60.Jd  
}

\begin{abstract}
The standard multidetector $\F$-statistic for continuous gravitational waves is susceptible to false
alarms from instrumental artifacts, for example monochromatic sinusoidal disturbances (``lines'').
This vulnerability to line artifacts arises because the $\F$-statistic compares the
signal hypothesis to a Gaussian-noise hypothesis, and hence is triggered by anything that resembles the signal
hypothesis \emph{more} than Gaussian noise.
Various ad-hoc veto methods to deal with such line artifacts have been proposed and used in the past.
Here we develop a Bayesian framework that includes an explicit alternative hypothesis to model disturbed data.
We introduce a simple line model that defines lines as signal candidates appearing only in one detector.
This allows us to explicitly compute the odds between the signal hypothesis and an extended noise hypothesis,
resulting in a new detection statistic that is more robust to instrumental artifacts.
We present and discuss results from Monte-Carlo tests on both simulated data and on
detector data from the fifth LIGO science run.
We find that the line-robust statistic retains the detection power of the
standard $\F$-statistic in Gaussian noise.
In the presence of line artifacts it is more sensitive, even compared to the popular $\F$-statistic
consistency veto, over which it improves by as much as a factor of two in detectable signal strength.
\end{abstract}

\maketitle

\section{Introduction}
\label{sec:introduction}

Spinning neutron stars with nonaxisymmetric deformations are expected to emit quasi-monochromatic and
long-lasting gravitational waves (GWs), commonly referred to as \emph{continuous waves}
(CWs)~\cite{prix06:_cw_review, bildsten1998:_gwns, ushomirsky2000:_deformations, JMcD2013:_maxelastic}.
One of the main search methods for CWs was developed for ground-based detectors (such as LIGO~\cite{LIGORef:2009},
Virgo~\cite{VirgoRef:2011}, GEO\,600~\cite{GEORef:2010}) and it is the so-called $\F$-statistic.
The $\F$-statistic was originally derived as a maximum-likelihood detection statistic
\cite{jks98:_data,cutler05:_gen_fstat}; it was later shown that it can also be derived as a Bayes
factor using somewhat unphysical  priors for the signal amplitude parameters \cite{prix09:_bstat}.

In the context of CW searches, the GW data is reasonably well described by an underlying Gaussian noise
distribution with additional non-Gaussian disturbances (see, e.g., Fig.~3 in
Ref.~\cite{abbott2004:_geoligo}, Fig.~3 in Ref.~\cite{aasi13:_eathS5} and Sec.~5.8 in
Ref.~\cite{behnke2013:_phdthesis}).
The $\F$-statistic corresponds to a binary hypothesis test between a signal hypothesis and a Gaussian-noise
hypothesis.
As a consequence, it is possible to obtain large $\F$-statistic values due to non-Gaussian disturbances in the
data, even if they are not well matched to the signal model. Large $\F$-statistic values only imply that the
signal hypothesis is a \emph{better} fit to the data than pure Gaussian noise, but they do not imply a
\emph{good} fit.

The most problematic instrumental artifacts for any specific analysis of GW data are typically
those that resemble the signal family it searches for, i.e., disturbances with non-negligible projection
onto the signal templates of a search.
For example, searches for short transient signals, such as bursts (e.g., from core-collapse
supernovae~\cite{fryer2011:_collapsereview}) or compact binary coalescences
(CBCs~\cite{abadie2010:_cbcrates}),
are most affected by ``glitches'' in the data, i.e., short broad-band
disturbances~\cite{Blackburn2008:_glitch, Aasi2012:_virgochar, Slutsky2010:_falsealarms,
prestegard2012:_transartifact}.

On the other hand, searches for CW signals are mainly affected by so-called ``lines'', i.e., narrow-band
disturbances that are present for a sizable fraction of the observation time.
Examples include the so-called \emph{mains} lines (i.e., lines at multiples of the 60\,Hz electrical
power system frequency for LIGO, or 50\,Hz for Virgo and GEO600), the resonance frequencies of the
detector suspensions (different for each detector), and lines from digital components -- see
Ref.~\cite{aasi13:_eathS5} for more details and a list of known instrumental lines identified in the data from
the fifth LIGO science run (S5), and Refs.~\cite{Aasi2012:_virgochar, Accadia2012:_noemi} for line
identification in Virgo data.

In this article, we apply a Bayesian model-selection approach using an additional alternative noise hypothesis
for lines.
Since the characteristics of the population of instrumental lines affecting CW searches are not well
understood, we use a very simple line model. The model is based on an observed
distinguishing feature of many lines, namely that they do not affect all detectors in the same way.
Hence, we define our line model as any feature in the data that resembles the signal template \emph{in only
one detector}.

This approach can also be seen as adding a coincidence criterion to the coherent multidetector
$\F$-statistic. Such a method is applicable only to multidetector CW searches, and in practice the most
recent $\F$-statistic-based searches have all used data from multiple detectors.
By employing this approach, we obtain a new line-robust detection statistic, generalizing the
$\F$-statistic.

The plan of this paper is as follows: In Sec.~\ref{sec:lit-review} we give a short review of existing
methods to deal with glitches and lines in GW data.
In Sec.~\ref{sec:hypotheses} we describe signal-noise hypotheses relevant for the detection problem at hand:
the standard Gaussian-noise hypothesis, in Sec.~\ref{sec:hypo_gauss}, the CW
signal hypothesis, in Sec.~\ref{sec:hypo_signal} and a new simple line hypothesis in Sec.~\ref{sec:hypo_line}.
In Sec.~\ref{sec:line-veto-stats-coh} we use these hypotheses to construct two new detection statistics, a
``line-veto'' statistic and a more general line-robust statistic.
We generalize the hypotheses and statistics to the case of semicoherent searches in
Sec.~\ref{sec:semicoherent}. Next we discuss the choice of prior parameters for the line-robust statistic and we
present a simple method, albeit somewhat ad hoc, to choose decent priors in Sec.~\ref{sec:tuning}. This
concludes the analytical part of the paper. In Sec.~\ref{sec:tests} we assess the performance of the new
statistics through a series of numeric tests: on fully synthetic data in Sec.~\ref{sec:tests_simdata} and on
LIGO S5 data in
Sec.~\ref{sec:tests_realdata}. We summarize our findings in Sec.~\ref{sec:conclusions} and give a short
outlook on applications and future generalizations of this approach.
The appendix~\ref{sec:expect-f-stat} contains a short derivation of the expected $\F$-statistic value under
the
simple line hypothesis.

\section{Existing methods to mitigate detector artifacts}
\label{sec:lit-review}

The problem of non-Gaussian artifacts in the data affects both searches for long-lived signals (e.g., the CWs
which are the topic of this paper) as well as searches for short-lived signals.
Short-lived signals are expected from the late phase of the inspiral of binaries of compact objects such as
neutron stars and black holes as well as from catastrophic events such as supernovae.
For short-lived signal searches the artifacts that are responsible for an increase in the false alarm rates
with respect to purely Gaussian noise manifest themselves as loud glitches in the time-domain data.
On the other hand, for long-lived signal searches the most troublesome artifacts are broadly speaking those
that appear in the Fourier spectra on the typical time scales of the search.

An interesting distinction in search pipelines is the order in which multidetector coherence and coincidence
are used. If the first step in the search is a coherent multidetector statistic (as
for the CW searches that we consider here), then the noise-artifact-mitigation strategy
may use subsequent consistency checks on the statistics from the individual detectors. If,
on the other hand, the first step was a single-detector search followed by a selection of triggers
where only coincidence in the various detectors is required, an additional multidetector coherent statistic
can serve as an artifact-mitigation technique.

A wide range of methods have been developed in order
to deal with instrumental artifacts. In the following sub-sections, we give a short review of
such methods for CBC, burst and CW searches.

Generally, we can distinguish between two fundamentally different approaches to artifact mitigation: Bayesian
model-selection and heuristic methods.
The former is based on \emph{explicit} alternative models whereas the latter consists in constructing
ad-hoc statistics to detect certain observed deviations from the GW signal model, which in the
Bayesian picture corresponds to a test against \emph{implicit} (and often unknown) alternative hypotheses.
A third ``hybrid'' approach uses Bayesian inference to directly construct empirical noise and
signal likelihoods \cite{cannon2008:_bayescoinc} using the actual data and simulated GW signals,
so-called \emph{injections}.

\subsection{Instrumental glitches in burst and CBC searches}
\label{sec:glitch-short-trans}

In searches for short-lived GW signals popular ad-hoc glitch-veto methods are the
$\chi^2$-veto \cite{allen2005:_chi2}, the null-stream veto \cite{wen2005:_nullstream}, and signal amplitude
consistency vetoes \cite{abbott2005:_cbcS2}.
These (among others) are commonly used in CBC searches (e.g., see Refs.~\cite{babak2013:_searchcbc,
harry2011:_targetcbc, abadie2012:_lowmasscbc}) and in burst searches (e.g., see
Refs.~\cite{sutton2010:_xpipeline, abbott2009:_s5burst, abadie2012:_allskyburst}).

For instance, in low-mass CBC searches the first step is a separate search in each detector.
After a cut on single-detector $\chi^2$ values, the glitch mitigation strategy consists in applying a
coincidence criterion and then
constructing on the surviving candidates a new multidetector statistic. This folds in the single-detector
statistics and $\chi^2$ values. Significance thresholds are set based on Monte-Carlo studies on actual data and
injections.

In searches for signals for which we lack a waveform model (i.e., generic bursts), a main multidetector
statistic is constructed that accounts appropriately for time delays and antenna responses of the different
detectors to the same putative GW. This statistic is then augmented by other statistics
(see Ref.~\cite{sutton2010:_xpipeline} for details) specifically designed to
further check for signal consistency across the detectors by means of appropriate veto conditions.

Various explicit glitch models have been considered, including Sine-Gaussians \cite{clark2007:_ringdown,
dalcanton2014:_sinegaussian} and wavelets \cite{littenberg2010:_artifacts}, and Bayesian approaches have also
been proposed to use these in constructing glitch-robust searches
\cite{clark2007:_ringdown, littenberg2010:_artifacts}. Notably, Veitch and Vecchio
\cite{veitch2010:_bayesian} have defined a glitch model describing coincident single-detector candidates with
independent amplitude parameters in different detectors.
On the other hand, the signal model requires candidates to be both coincident and coherent across all
detectors.
Both hypotheses would fit a true signal equally well, but the glitch hypothesis would be weighed down by its
larger prior volume (``Occam's razor''). In the case of glitches, however, the glitch hypothesis will
generally provide a much better fit, allowing it to overcome its larger prior volume.

\subsection{Instrumental lines in CW searches}
\label{sec:deal-with-instr}

The most commonly used approaches to deal with instrumental lines in CW searches are all heuristic and can be
summarized as follows:
\begin{enumerate}[(i)] \itemsep1pt \parskip0pt
\item \emph{Line cleaning} This is a widely used approach with many variants. It consists in effectively
  excluding frequency bands from the search when they are known or believed to be affected by instrumental
  lines.
  This could be either as a result of previous detector characterization work or because the frequency-domain
  data was flagged as particularly disturbed (referred to as {\emph{line flagging}}).
  Among the examples of this approach are the LIGO/Virgo searches
  \cite{aasi13:_eathS5, abadie12:_powerflux,aasi2013:_gc-search, abbot2008:_s4cw, abadie2010:_casa}.

  A downside of this method is the relatively large fraction of the total frequency band it
  typically vetoes. For example, in Ref.~\cite{abadie12:_powerflux} it vetoed a total of 270\,Hz out of the
  1140\,Hz searched, i.e., $\sim24\%$ of the data.
  Furthermore, this method is either limited to known instrumental lines or, when the line-flagging variant
  is used, its efficacy is limited to strong  disturbances.
  Weaker disturbances can only be identified with time baselines much longer than the ones typically
  used by the line-flagging algorithms. Furthermore, the Fourier-transform-based line-flagging algorithm is not
  optimally suited to detect lines with nonconstant frequency.

\item \emph{S-veto} This is a method to remove candidates from a (frequency and spin-down dependent) region of
  the sky.
  This region is typically around the poles, where the corresponding signal templates are not well
  distinguished from typical instrumental line artifacts. This method was initially developed in PowerFlux
  \cite{abbot2008:_s4cw} and subsequently adapted to $\F$-statistic searches \cite{abbott2009:_s4eath}.

  The fraction of the total parameter space vetoed a priori through this approach can again be
  quite large, for example about $\sim 30\%$ in Ref.~\cite{abbott2009:_s4eath}.

\item \emph{$\F$-statistic consistency veto} If a candidate from a multidetector search has a
  single-detector $\F$-statistic value exceeding its multidetector $\F$-statistic, then it is
  vetoed as a likely instrumental line. This approach was described in more detail and tested in
  Ref.~\cite{aasi13:_eathS5} and Refs.~\cite{behnke2013:_phdthesis, aasi2013:_gc-search}.
\end{enumerate}

The approach proposed in the present work is not a heuristic method as the ones described above. Instead, it
shares some similarities to the glitch-robust method proposed in Ref.~\cite{veitch2010:_bayesian}, but it
differs in the following:
In the incoherent CBC pipeline, any candidate is already required to be \emph{coincident} between detectors,
so the method of Ref.~\cite{veitch2010:_bayesian} adds the requirement of multidetector \emph{coherence} to
distinguish GW signals from glitches.
In our case, instead, we start from the \emph{coherent} multidetector $\F$-statistic and add a
\emph{coincidence} requirement to distinguish CW signals from (noncoincident) lines.

Currently we do not include coincident lines in the alternative hypothesis, as we expect that this would
substantially weaken the detection power of this method.
More work is required to deal with coincident lines that trigger the same templates in multiple detectors.

However, the prevalence of coincident lines in detector data appears to be limited.
For example, the lines of known instrumental origin identified in the LIGO S5 data from the two detectors (see
Tables VI and VII in Ref.~\cite{aasi13:_eathS5}) overlap by 1.6~Hz, corresponding to about 11\% of the
contaminated bandwidth.
Furthermore, in the $\F$-statistic based analysis~\cite{aasi13:_eathS5}, 0.46\% of final high-significance
candidates passed the $\F$-statistic consistency veto and therefore could be considered as caused by
coincident lines.

The approach taken here is that in a full CW search pipeline the noncoincident line model would serve as a
cheap and simple ``first line of defense'' to reduce the number of spurious candidates, while more
sophisticated steps can be applied to the surviving candidates in later steps.

\section{Hypotheses about the observed data}
\label{sec:hypotheses}

Let $x^X(t)$ be the time series of GW strain measured in a detector $X$, where we use
$X,Y,\ldots$ as detector indices. Following the multidetector notation from
Refs.~\cite{cutler05:_gen_fstat,prix06:_searc}, using boldface indicates a multidetector vector, i.e.,
we write $\dVx(t)$ for the multidetector data vector with components $x^X(t)$.

We will consider three different
hypotheses about the observed data $\dVx$ and derive their posterior probabilities: the Gaussian noise
hypothesis $\HypG$, the CW signal hypothesis $\HypS$ and a simple ``line'' hypothesis $\HypL$.

\subsection{The Gaussian noise hypothesis \texorpdfstring{$\HypG$}{HG}}
\label{sec:hypo_gauss}

The Gaussian-noise hypothesis $\HypG$ states that the measured multidetector time series $\dVx(t)$ only
contains stationary Gaussian noise, which we denote as $\dVn(t)$, i.e.,
\begin{equation}
  \label{eq:hypG}
  \HypG: \dVx(t) = \dVn(t)\,,
\end{equation}
with a single-sided power-spectral density (PSD) $\detVec{S}_n$ that is assumed to be known.
The corresponding likelihood for measuring the data $\dVx$ can therefore be written
as
\begin{equation}
  \label{eq:gaussian}
  \prob{\dVx}{\HypG} = \kappa\,\eto{-\frac{1}{2}\scalar{\dVx}{\dVx}}\,,
\end{equation}
where $\kappa$ is a data-independent normalization constant, and the scalar product is defined as
\begin{equation}
  \label{eq:scalarproduct}
  \scalar{\dVx}{\dVy} \equiv \sum_X \frac{1}{\SnX}\int_0^T x^X(t)\,y^X(t)\,dt\,,
\end{equation}
assuming that the noise spectra $\SnX$ are uncorrelated between different detectors $X$ and constant over the
(narrow) frequency band of interest.
For simplicity of notation we omit the sometimes customary notation of a conditional ``$I$'' denoting all
implicit and explicit model assumptions, i.e., we write $\prob{a}{b}$ as a shortcut for $\prob{a}{b\,,I}$, and
$\probI{a}$ as an abbreviation for $\prob{a}{I}$.

The posterior probability for $\HypG$ given the observed data $\dVx$ follows from Bayes' theorem as
\begin{equation}
  \label{eq:pHG}
  \prob{\HypG}{\dVx} = \frac{\probI{\HypG}}{\probI{\dVx}}\,\kappa\,\eto{-\frac{1}{2}\scalar{\dVx}{\dVx}} \,,
\end{equation}
where $\probI{\HypG}$ is the prior probability for the Gaussian-noise hypothesis.
The normalization $\probI{\dVx}$ depends on the full set of assumed hypotheses $\{\Hyp_i\}$, i.e.,
\mbox{$\probI{\dVx} = \sum_i \prob{\dVx}{\Hyp_i}\,\probI{\Hyp_i}$}, but in the following we will only consider
the \emph{odds} between different hypotheses, where this term drops out.

\subsection{The CW signal hypothesis \texorpdfstring{$\HypS$}{HS}}
\label{sec:hypo_signal}

The hypothesis $\HypS$ for CW signals \cite{jks98:_data,prix06:_cw_review} states that the data $\dVx$
contains a CW signal $\dVh$ in addition to Gaussian noise $\dVn$, namely $\dVx = \dVn + \dVh$.

The signal model $\dVh$ depends on a number of (generally unknown) signal parameters. For
practical reasons, we usually distinguish between the set of four \emph{amplitude parameters} $\Amp$ and the
remaining \emph{phase-evolution parameters} $\Dop$, i.e., we write the CW signal family as $\dVh(t;\Amp,\Dop)$.

To fully specify the signal hypothesis, we therefore need a prior probability distribution
$\prob{\Amp,\Dop}{\HypS}$ for the signal parameters, i.e.,
\begin{equation}
  \begin{split}
  \label{eq:hypS}
  \HypS: \dVx(t) = \dVn(t) + \dVh(t;\Amp,\Dop)\\
  \text{with prior}\;\;\prob{\Amp,\Dop}{\HypS}\,.
\end{split}
\end{equation}

The amplitude parameters $\Amp$ describe the signal amplitude $h_0$, the inclination angle $\iota$, the
polarization angle $\psi$ and the initial phase $\phio$. As first shown in Ref.~\cite{jks98:_data}, a
particular parametrization $\Amp^\mu = \Amp^\mu ( h_0,\cosi,\psi,\phio)$, with $\mu = 1\ldots4$, allows one to
write the signal model in the factorized form
\begin{equation}
  \label{eq:Amuhmu}
  \dVh(t;\Amp,\Dop) = \Amp^\mu\,\dVh_\mu(t;\Dop)\,,
\end{equation}
in terms of four basis functions $\dVh_\mu(t;\Dop)$ and using the automatic summation convention over repeated
indices.

In order to simplify the following discussion and notation, we follow the approach
of Refs.~\cite{prix09:_bstat,prix11:_transient} and formally restrict ourselves to a single-template statistic
in $\Dop$. This is equivalent to the assumption of known phase parameters, i.e., $\Dop = \Dop_\sig$. This can
be done without loss of generality, as for unknown $\Dop\in\DopSpace$ this analysis would apply for each
template $\Dop_i\in\DopSpace$, and one would then marginalize over the prior parameter space $\DopSpace$.
Studying this in further detail is outside the scope of the present work.
We will therefore assume a prior of the form
\begin{equation}
  \label{eq:priorAlambda}
  \prob{\Amp,\Dop}{\HypS} = \prob{\Amp}{\HypS}\,\delta(\Dop - \Dop_\sig)\,,
\end{equation}
and drop the phase-evolution parameters $\Dop$ from the following expressions.

We can obtain the likelihood for \emph{a particular} signal $\dVh(t;\Amp)$ by noting
that, according to $\HypS$, the combination $\left[\dVx - \dVh(\Amp)\right]$ is described by Gaussian noise.
In fact, by inserting the signal factorization from Eq.~\eqref{eq:Amuhmu} and by factoring out terms
equivalent to the Gaussian noise likelihood from Eq.~\eqref{eq:pHG}, we obtain
\begin{align}
  \prob{\dVx}{\HypS,\Amp}  &= \kappa \, \eto{-\frac{1}{2} \scalar{\dVx-\dVh(\Amp)}{\dVx-\dVh(\Amp)}} \notag\\
                           &= \kappa \, \eto{-\frac{1}{2} \scalar{\dVx}{\dVx}} \,
                              \eto{\scalar{\dVx}{\Amp^\mu\dVh_\mu} - \frac{1}{2}
                              \scalar{\Amp^\mu\dVh_\mu}{\Amp^\nu\dVh_\nu}}   \label{eq:likeli_HSA}\\
                           &= \prob{\dVx}{\HypG} \exp\left[\Amp^\mu \, x_\mu - \frac{1}{2} \Amp^\mu
                             \M_{\mu\nu}\Amp^\nu\right],\notag
\end{align}
where we introduced the four projections $x_\mu$ of the data and the (symmetric positive-definite) matrix
$\M_{\mu\nu}$ as
\begin{equation}
  \label{eq:xmuMmunu}
  x_\mu \equiv \scalar{\dVx}{\dVh_\mu} \quad \text{and} \quad
  \M_{\mu\nu} \equiv \scalar{\dVh_\mu}{\dVh_\nu}\,.
\end{equation}

The \emph{marginal} likelihood $\prob{\dVx}{\HypS}$ (sometimes referred to as ``evidence'') for the
signal hypothesis from Eq.~\eqref{eq:hypS} can be obtained by marginalizing over the unknown
amplitudes $\Amp$, namely
\begin{equation}
 \prob{\dVx}{\HypS} = \int \prob{\dVx}{\HypS,\Amp}\,\prob{\Amp}{\HypS}\,d\Amp\,. \label{eq:likeli_HSmarg}
\end{equation}
This integral can be solved analytically for certain choices of amplitude priors $\prob{\Amp}{\HypS}$.
In particular, as discussed in Refs.~\cite{prix09:_bstat,prix11:_transient}, for the (somewhat unphysical)
prior that is uniform in $\Amp^\mu$, we can recover the standard $\F$-statistic, namely, assuming
\begin{equation}
  \label{eq:priorA}
  \prob{\{\Amp^\mu\}}{\HypS} = \left\{\begin{array}{ll}
      C & \text{for} \quad h_0^4(\Amp) < \frac{70\,\cF}{\sqrt{|\M|}}\,.\\
      0 & \text{otherwise}\,,
    \end{array}\right.
\end{equation}
where $|\M|$ is the determinant of $\M_{\mu\nu}$ and $\cF$ is an ad-hoc cutoff\footnote{This
translates to the notation of Ref.~\cite{prix11:_transient} via $\cF = \frac{\rhomax^4}{70}$.}
used to normalize the prior, namely, \mbox{$C = \frac{\sqrt{|\M|}}{(2\pi)^2}\, \cF^{-1}$}.

Using this prior and taking the integration boundary to infinity, $\cF\rightarrow\infty$, we obtain
the (marginal) signal likelihood, from Eq.~\eqref{eq:likeli_HSmarg}, in the form
\begin{equation}
  \label{eq:likeli_HS}
  \prob{\dVx}{\HypS} = \prob{\dVx}{\HypG} \, \cF^{-1}\,\eto{\F(\dVx)}\,,
\end{equation}
where we define the (coherent) multidetector $\F$-statistic as
\begin{equation}
  \label{eq:Fstat}
  2\F(\dVx) \equiv x_\mu\,\M^{\mu\nu}\,x_\nu\,,
\end{equation}
and $\M^{\mu\nu}$ denotes the inverse matrix to $\M_{\mu\nu}$, i.e.,
$\M_{\mu\alpha}\M^{\alpha\nu} = \delta_{\mu}^{\nu}$.
We obtain the posterior probability for the signal hypothesis as
\begin{equation}
  \label{eq:pHS_final}
  \prob{\HypS}{\dVx} = \oSG\,\cF^{-1}\,\prob{\HypG}{\dVx}\,\eto{\F(\dVx)}\,,
\end{equation}
where $\oSG\equiv \probI{\HypS}/\probI{\HypG}$ denotes the prior odds between the signal- and
Gaussian-noise hypotheses.

The posterior odds between signal hypothesis $\HypS$ and Gaussian-noise hypothesis $\HypG$ are therefore
equivalent to the standard multidetector $\F$-statistic\footnote{\emph{Equivalence} in the Neyman-Pearson
sense: the same false-dismissal as a function of false-alarm probability.}, as we see by writing
\begin{equation}
  \label{eq:OSG}
  \OSG(\dVx) \equiv \frac{\prob{\HypS}{\dVx}}{\prob{\HypG}{\dVx}}
                  = \oSG\,\cF^{-1}\, \eto{\F(\dVx)}\,.
\end{equation}
Note that the corresponding (marginal) likelihood ratio
\begin{equation}
  \label{eq:2}
  \BSG(\dVx) \equiv \frac{\prob{\dVx}{\HypS}}{\prob{\dVx}{\HypG}} = \cF^{-1}\, \eto{\F(\dVx)}\,,
\end{equation}
is generally known as the \emph{Bayes factor}, and is closely related to the odds via $\OSG(\dVx) = \oSG\,\BSG(\dVx)$.

While this statistic is close to optimal for detecting signals in pure Gaussian noise \cite{prix09:_bstat}, it
is vulnerable to various signal-like instrumental artifacts in the data.
As discussed in Sec.~\ref{sec:introduction}, we see from Eqs.~\eqref{eq:OSG} and \eqref{eq:2} that detector
artifacts can trigger $\OSG(\dVx)$ or $\BSG(\dVx)$, provided they resemble $\HypS$ \emph{more} than $\HypG$
even if the agreement with $\HypS$ is poor.
In order to deal with this problem, we need to introduce an alternative hypothesis, which describes
instrumental lines \emph{better} than $\HypS$.

\subsection{Simple line hypothesis: A CW-like disturbance in a single detector}
\label{sec:hypo_line}

Here we introduce a simple line hypothesis designed to match one prominent feature of many instrumental lines,
distinguishing them from CW signals: the fact that they appear only in one detector.
Inspired by this, we reuse the signal hypothesis from Eq.~\eqref{eq:hypS} in order to define a line in
detector $X$ :
\begin{equation}
  \begin{split}
  \label{eq:hypLX}
  \HypL^{X} :  x^{X}(t) = n^{X}(t) + h^{X}(t;\Amp^{X})\\
  \text{with prior}\quad\prob{\Amp^{X}}{\HypL^{X}}\,.
\end{split}
\end{equation}
We would expect lines to have a different amplitude distribution than real signals, but in the absence of any
more detailed knowledge on this point, we choose to reuse the signal amplitude prior given by
Eq.~\eqref{eq:priorA} for $\prob{\Amp^X}{\HypL^X}$. This choice simplifies the following
calculations. In analogy to Eq.~\eqref{eq:pHS_final}, we directly obtain the probability for $\HypL^X$:
\begin{equation}
  \label{eq:pHLX}
  \prob{\HypL^X}{x^X} = \cF^{-1}\,\prob{\HypG^X}{x^X}\,\oLG^X\,\eto{\FX(x^X)}\,.
\end{equation}
Here we define the per-detector prior line odds \mbox{$\oLG^X \equiv {\probI{\HypL^X}}/{\probI{\HypG^X}}$},
which encode prior knowledge about how likely a line is, compared to pure Gaussian noise, in a given
template $\Dop$ and detector $X$. The detector-specific $\F$-statistic $\FX(x^X)$ is simply given by
Eq.~\eqref{eq:Fstat} restricted to detector $X$.

For multiple detectors we can now formulate the simple line hypothesis $\HypL$ as a CW-like disturbance
$\HypL^{X}$ in any one detector $X$ and data consistent with Gaussian noise $\HypG^Y$ in all other detectors
$Y\not=X$:
\begin{equation}
  \begin{split}
    \HypL \equiv& \left( \HypL^1 \AND \HypG^2 \AND \HypG^3 \ldots\right) \OR \\
    & \left( \HypG^1 \AND \HypL^2 \AND \HypG^3 \ldots \right) \OR \ldots\,. \label{eq:hypL}
  \end{split}
\end{equation}
Note that in this approach $\HypL$ does not include lines that are coincident across different detectors,
which is postponed to future work.

We assume the different detectors to be independent to the extent that
knowing $\HypG^X$ or $\HypL^X$ for detector $X$
does not inform us about $\HypG^Y$ or $\HypL^Y$ for other detectors $Y\not=X$. We also assume the different
alternatives in Eq.~\eqref{eq:hypL} to
be mutually exclusive. The laws of probability therefore yield
\begin{align}
  \prob{\HypL}{\dVx} &= \prob{\HypL^1}{x^1} \prob{\HypG^2}{x^2} \prob{\HypG^3}{x^3}\times \ldots \nonumber \\
  & + \prob{\HypG^1}{x^1} \prob{\HypL^2}{x^2} \prob{\HypG^3}{x^3}\times \ldots \nonumber \\
  & + \ldots \nonumber \\
  &= \sum_{X} \prob{\HypL^{X}}{x^{X}} \prod_{Y\not=X} \prob{\HypG^Y}{x^Y}\,. \label{eq:pHL_initial}
\end{align}

By combining Eqs.~\eqref{eq:hypL}, \eqref{eq:pHLX} and the (per-detector) Gaussian-noise probability from
Eq.~\eqref{eq:pHG}, we find the posterior probability for the line hypothesis $\HypL$ as
\begin{equation}
  \label{eq:pHL_inserted}
  \prob{\HypL}{\dVx} = \cF^{-1}\,\prob{\HypG}{\dVx} \, \sum_X \oLG^X\,\eto{\FX(x^X)}\,,
\end{equation}
where we used the fact that $\prod_X \prob{\HypG^X}{x^X} = \prob{\HypG}{\dVx}$. Note that
\begin{equation}
  \label{eq:sumlX}
  \sum_X \oLG^X = \frac{\probI{\HypL}}{\probI{\HypG}} \equiv \oLG\,,
\end{equation}
where $\oLG$ denotes the prior odds for a line versus Gaussian noise (in the present template $\Dop$)
including all detectors.

It will be convenient to define relative detector weights $\rX$ for the prior line odds, namely for
$\Ndet$ detectors:
\begin{equation}
  \label{eq:rX}
  \rX \equiv \frac{\oLG^X}{\oLG / \Ndet}\,,\quad\text{such that}\quad \sum_X \rX = \Ndet\,.
\end{equation}
If all detectors are equally likely to contain a line, then $\rX = 1$ for all $X$.
We further denote the average of a quantity $Q^X$ over detectors as
\begin{equation}
  \label{eq:1}
  \avgX{Q^X} \equiv \frac{1}{\Ndet} \sum_X Q^X\,,
\end{equation}
and hence $\avgX{\rX} = 1$.
By using these definitions, we can write Eq.~\eqref{eq:pHL_inserted} as follows:
\begin{equation}
  \label{eq:pHL_avg}
  \prob{\HypL}{\dVx} = \cF^{-1} \, \prob{\HypG}{\dVx} \, \oLG\, \avgX{\rX\,\eto{\FX(x^X)}}\,.
\end{equation}

\section{Coherent line-robust statistics}
\label{sec:line-veto-stats-coh}

We use the posterior line probability of Eq.~\eqref{eq:pHL_avg} to compute the odds for
additional model comparisons, thereby extending the standard multidetector $\F$-statistic
given by Eq.~\eqref{eq:OSG}.
In particular, we consider two approaches:
\begin{enumerate}[(i)] \itemsep1pt \parskip0pt
\item Define a ``line-veto'' statistic as the odds between the signal hypothesis $\HypS$ and
  the line hypothesis $\HypL$.
  This may be useful, for example, as a follow-up statistic for strong candidates from
  an initial $\F$-statistic search, which compared $\HypS$ versus Gaussian noise $\HypG$.
  In such a two-stage approach, one would test the signal hypothesis against the line-hypothesis
  if the Gaussian-noise hypothesis has been ruled out with sufficient confidence.

\item \emph{Extend} the standard signal-versus-Gaussian-noise odds $\OSG(\dVx)$ to a more
  line-robust statistic $\OSN(\dVx)$ by allowing the noise hypothesis to include either pure Gaussian
  noise $\HypG$ or a line $\HypL$.
\end{enumerate}

\subsection{Line-veto statistic \texorpdfstring{$\OSL(\dVx)$}{OSL(x)}}
\label{sec:line-veto-stat}

Using the posterior probabilities given by Eqs.~\eqref{eq:pHS_final} and \eqref{eq:pHL_avg}, we obtain the
posterior signal-versus-line odds as
\begin{equation}
  \label{eq:OSL}
  \OSL(\dVx) \equiv \frac{\prob{\HypS}{\dVx}}{\prob{\HypL}{\dVx}} =
  \oSL \; \frac{ \eto{\F(\dVx)} }{\avgX{\rX\,\eto{\FX(x^X)}}}\,,
\end{equation}
with the prior odds $\oSL \equiv \probI{\HypS}/\probI{\HypL} = \oSG/\oLG$. Note
that the amplitude-prior cutoff $\cF$ has disappeared, as we have used the same amplitude
prior on lines and signals.

In the following we will often neglect the dependency on $\dVx$ and $x^X$ to simplify notation.
It is instructive to consider the log-odds, which we can write as
\begin{align}
 \ln\OSL &= \ln \oSL + \F - \ln \avgX{\rX\eto{\FX}} \nonumber \\
               &= \ln \oSL + \F - \Fmaxp - \ln \avgX{\rX\eto{\left(\FX -
                  \Fmaxp\right)}}\,,\label{eq:logOSL}
\end{align}
where we define
\begin{equation}
  \label{eq:Fmaxp}
  \Fmaxp \equiv \max_X \left( \FX + \ln \rX \right) \,.
\end{equation}
The terms in the detector-average in Eq.~\eqref{eq:logOSL} are bounded within $[0, 1]$, with at least one term
being equal to $1$.
Hence, the logarithmic average $\ln\avgX{\ldots}$ is bounded within
$[-\ln\Ndet,0]$, i.e., of order 1.

Actually, for strong $\F$-statistic candidates, i.e., $\F\gg1$, the logarithmic correction is
negligible, and therefore we can approximate
\begin{equation}
  \label{eq:logOSL_approx}
  \ln\OSL(\dVx) \approx \F(\dVx) - \Fmaxp(\dVx) + \ln\oSL \,.
\end{equation}
Without prior knowledge about one detector being more affected by instrumental lines than others,
we would have $\rX = 1$ and therefore $\Fmaxp(x) = \max_X \FX(x^X)$.
Considered as a detection statistic, $\OSL(\dVx)$ is therefore approximately equivalent to the
difference between the multidetector $\F$-statistic and the largest $\F$-statistic value from the
individual detectors.

By choosing a special threshold of $\OSL(\dVx)=\oSL$ and assuming equal prior line probabilities for all
detectors, we recover the well-known \emph{$\F$-statistic consistency veto}, namely
\begin{equation}
  \label{eq:F-veto}
  \textrm{If}\quad \F(\dVx) < \max_{X}\{\F^X(x)\}\;\;\implies\;\textrm{veto the candidate}\,,
\end{equation}
which has been successfully used and tested in
Refs.~\cite{aasi13:_eathS5, behnke2013:_phdthesis, aasi2013:_gc-search}.
Combining this veto with $\F$-statistic ranking corresponds to defining a new statistic:
\begin{equation}
  \label{eq:Fveto}
  \Fveto(\dVx) \equiv \left\{
    \begin{array}{cc}
      \F(\dVx)\quad & \textrm{if } \F(\dVx) \ge \max_{X}\{\F^X(x)\}\,,\\
      0        & \textrm{otherwise}\,,
    \end{array}
  \right.
\end{equation}
which we will refer to as the \FvetoName.

\subsection{Line-robust detection statistic \texorpdfstring{$\OSN(\dVx)$}{OSN(x)}}
\label{sec:extend-detect-stat}

From the standpoint of probability theory it is more natural to use the line hypothesis to extend
what we mean by ``noise'', namely, either pure Gaussian noise $\HypG$ or a line $\HypL$.
Hence, we introduce an extended noise hypothesis as
\begin{equation}
  \label{eq:hypN}
  \HypN : \left( \HypG \OR \HypL \right) \,.
\end{equation}
Since we take $\HypG$ and $\HypL$ to be mutually exclusive, the posterior probability for $\HypN$ is
\begin{align}
  \prob{\HypN}{\dVx} &= \prob{\HypG}{\dVx} + \prob{\HypL}{\dVx} \nonumber \\
                     &= \prob{\HypG}{\dVx} \left( 1 + \cF^{-1} \,\oLG \,\avgX{\rX \eto{\FX(x^X)}} \right) \,,
\label{eq:pHN}
\end{align}
where we have used Eq.~\eqref{eq:pHL_avg} for the explicit line posterior.

Interestingly, we can express the odds $\OSN(\dVx)$ of the signal versus extended noise hypotheses as
\begin{equation}
  \label{eq:OSN_initial}
  \OSN(\dVx) \equiv \frac{\prob{\HypS}{\dVx}}{\prob{\HypN}{\dVx}}
  = \left[ \OSG^{-1}(\dVx) + \OSL^{-1}(\dVx) \right]^{-1}\,.
\end{equation}
We can compare this result with the ad-hoc two-stage approach discussed previously, where one would
set two independent thresholds on $\OSG$ and on $\OSL$. As we see from Eq.~\eqref{eq:OSN_initial}, instead the
laws of probability tell us to compute the harmonic sum of $\OSG$ and $\OSL$ and to set a single threshold on
the resulting statistic.

Inserting the explicit expressions provided by Eqs.~\eqref{eq:pHN} and \eqref{eq:pHS_final}, we obtain
\begin{equation}
  \label{eq:OSN_cstar}
  \OSN(\dVx) = \frac{\oSG \, \eto{\F(\dVx)}}
  {\cF + \oLG \avgX{\rX \eto{\FX}}}\,.
\end{equation}
The amplitude-prior cutoff parameter $\cF$ from Eq.~\eqref{eq:priorA} is only a scale
factor in $\OSG$ and thus not relevant for the performance as a detection statistic, and it is canceled
out completely in $\OSL$.
However, in $\OSN$ this parameter does affect the properties of the resulting statistic.

We can rewrite Eq.~\eqref{eq:OSN_cstar} by introducing prior odds $\oSN \equiv \probI{\HypS}/\probI{\HypN}$ and
[noting that \mbox{$\oSG = \oSN\left( 1 + \oLG\right)$}] we obtain
\begin{equation}
  \label{eq:OSN_final}
  \OSN(\dVx) = \oSN \, \frac{\eto{\F(\dVx)}}
  {(1-\lineprob)\,\eto{\Ftho} + \lineprob\, \avgX{\rX \eto{\FX(x^X)}}}\,,
\end{equation}
where we define the prior line probability $\lineprob$ as
\begin{equation}
  \label{eq:lineprob}
  \lineprob \equiv \frac{\oLG}{1 + \oLG} = \frac{\probI{\HypL}}{\probI{\HypN}} = \prob{\HypL}{\HypN}\in [0,1]\,,
\end{equation}
and we used a more natural reparametrization of $\cF$ by defining
\begin{equation}
  \label{eq:Fth0}
  \Ftho \equiv \ln \cF\,.
\end{equation}

\subsubsection{Limiting cases of \texorpdfstring{$\OSN(\dVx)$}{OSN}}
\label{sec:limiting-behavior}

We now consider the limiting behavior of $\OSN$ as a function of the line prior $\lineprob$ and of the
single-detector $\FX(x)$ values.
We see from Eq.~\eqref{eq:OSN_final} that $\OSN(\dVx)$ reduces to the $\F$-statistic if we are certain that
there are no lines, i.e., $\OSN(\dVx) \rightarrow\OSG(\dVx)\propto\eto{\F(\dVx)}$ for $\lineprob\rightarrow0$.
On the other hand, it reduces to the pure line-veto statistic of Eq.~\ref{eq:OSL} when we believe the noise to
be completely dominated by lines, i.e., $\OSN(\dVx) \rightarrow \OSL(\dVx)$ for $\lineprob \rightarrow 1$.

For fixed $\lineprob$ we see that the transition between these two extremes depends on
the $\FX(x)$ values compared to the prior scale $\Ftho$. To illustrate this more clearly, we first rewrite
Eq.~\eqref{eq:OSN_final} using the relations $\oSN=\lineprob\,\,\oSL$ and $(1-\lineprob)/\lineprob=\oLG^{-1}$.
Introducing the ``transition scale'' $\Fth$ as
\begin{equation}
  \label{eq:Fth}
  \Fth \equiv \Ftho - \ln {\oLG}\,,
\end{equation}
we obtain
\begin{equation}
  \label{eq:OSN_Fstar}
  \OSN(\dVx) = \oSL\,\frac{\eto{\F(\dVx)}}{\eto{\Fth} + \avgX{\rX \eto{\FX(x^X)}}}\,.
\end{equation}
From this reparametrization, we see that $\Fth$ defines the scale of a smooth transition of $\OSN(\dVx)$
between $\OSG(\dVx)\propto\eto{\F(\dVx)}$ and $\OSL(\dVx)$ depending on the values of $\FX$: namely the
``line-veto term'' $\avgX{\rX \eto{\FX}}$ in Eq.~\eqref{eq:OSN_Fstar} only starts to play a role when it is
comparable to $\eto{\Fth}$.

To see this more explicitly, we write the log-odds as
\begin{equation}
  \begin{split}
  \label{eq:logOSN}
  \ln &\OSN(\dVx) = \ln \oSL + \F(\dVx) - \Fmaxpp(x)\\
   & - \ln \left( \eto{\Fth-\Fmaxpp} + \avgX{\rX \eto{\FX(x^X)-\Fmaxpp(\dVx)} } \right)\,,
  \end{split}
\end{equation}
where we define $\Fmaxpp(\dVx) \equiv \max\left( \Fth,\, \FX(x^X) + \ln\rX\right)$.
The logarithmic correction is of order unity, therefore this effectively corresponds to
$\ln\OSL(\dVx)$ when \mbox{$\max( \FX(x) + \ln \rX ) > \Fth$}, and to $\ln\OSG(\dVx)$ otherwise.

In practice it can be difficult to determine good prior values for $\Ftho$, due to the unphysical choice of
amplitude priors in Eq.~\eqref{eq:priorA}. We will discuss this issue in more detail in
Sec.~\ref{sec:choosing-prior-value}.

This transitioning behavior is reminiscent of the two-stage line-veto approach discussed in
Sec.~\ref{sec:line-veto-stats-coh}. There one applies a line veto only to candidates that are ``strong'' in
terms of $\OSG(\dVx)\propto\eto{\F(\dVx)}$, which means that the Gaussian-noise hypothesis
is already considered sufficiently unlikely.
Note, however, that for $\OSN(\dVx)$ the transition from $\OSG(\dVx)$ to $\OSL(\dVx)$ is smooth and depends on
the strength of the single-detector statistics $\FX(x^X)$ rather than the multidetector statistic $\F(\dVx)$.

\section{Semicoherent line-robust statistics}
\label{sec:semicoherent}

For unknown signal parameters $\Dop$, the use of the fully coherent (in time) $\F$-statistic is usually
prohibitive in terms of computing cost.
Thus, \emph{semicoherent} methods are typically used, being more sensitive at fixed computing cost
\cite{brady2000:_hierarchical,prix12:_optimal}.
In this approach the data $\dVx$ is divided into $\Nseg$ segments of shorter duration, denoted as
$\{\dVx_\segk\}_{k=1}^{\Nseg}$. The coherent statistic $\F_\segk(\dVx_\segk;\Dop)$ in a template $\Dop$ is
computed for each segment $\segk$ separately and then combined \emph{incoherently}, typically by summing over
all data segments. This is often referred to as the ``StackSlide'' method. Other incoherent combinations
such as the ``Hough transform'' method \cite{krishnan04:_hough} will not be discussed here.
The following discussion refers to the statistic in a single template $\Dop$, and we will therefore simplify
the notation again by dropping $\Dop$.

As shown in Ref.~\cite{prix11:_transient}, the semicoherent StackSlide $\F$-statistic can be derived by
relaxing the requirement of consistent signal amplitudes $\Amp$ across different segments, i.e., allowing for
a set of $\Nseg$ independent amplitude parameters $\Amp_\segk$ in Eq.~\eqref{eq:hypS}. This defines the
semicoherent signal hypothesis $\HypSsc$ as
\begin{equation}
  \label{eq:hypSsc}
  \HypSsc : \dVx_\segk = \dVn_\segk + \Amp_\segk^\mu\,\dVh_\mu\,,\quad
  \text{for } k = 1, \ldots \Nseg\,,
\end{equation}
where here and in the following the hat $\sc{\;\;}$ notation refers to semicoherent quantities.

For the per-segment amplitude priors $\prob{\Amp_\segk}{\HypSsc}$, we reuse the amplitude prior given by
Eq.~\eqref{eq:priorA}. Hence, by marginalization as in Eq.~\eqref{eq:likeli_HSmarg}, we obtain
the posterior
\begin{equation}
  \label{eq:pHSsc}
  \prob{\HypSsc}{\dVx} = \oSGsc\,\prob{\HypG}{\dVx}\, \cF^{-\Nseg}\, \eto{\scF(\dVx)}\,,
\end{equation}
where we define the StackSlide $\F$-statistic $\scF$ in parameter-space point $\Dop$ as
\begin{equation}
  \label{eq:avFstat}
  \scF(\dVx;\Dop) \equiv \sum\limits_{k=1}^{\Nseg} \F_\segk(\dVx_\segk;\Dop)\,.
\end{equation}
For Gaussian noise we have $\HypGsc=\HypG$, but for consistency of notation we still write $\HypGsc$ throughout
this section.
The posterior odds between the signal and Gaussian-noise hypotheses across the $\Nseg$ segments is
\begin{equation}
  \label{eq:OSGsc}
  \OSGsc(\dVx) \equiv \frac{ \prob{\HypSsc}{\dVx} }{ \prob{\HypGsc}{\dVx} } =
  \oSGsc\,\cF^{-\Nseg}\, \eto{\scF(\dVx)}\,.
\end{equation}

We can now generalize the single-detector line hypothesis of Eq.~\eqref{eq:hypL} to the semicoherent case as
was done for the signal hypothesis in Eq.~\eqref{eq:hypSsc}, namely,
\begin{equation}
  \begin{split}
  \label{eq:hypLsc}
  \HypLsc =& \left(\HypLsc^{1} \AND \HypGsc^{2} \AND \HypGsc^{3}\ldots\right) \OR\\
  &\left(\HypGsc^{1} \AND \HypLsc^{2} \AND \HypGsc^{3} \ldots\right) \OR \ldots
  \end{split}
\end{equation}
The probability of the line hypothesis in detector $X$ across all segments is
\begin{equation}
  \label{eq:pHLXsc}
  \prob{\HypLsc^X}{x^X} = \prob{\HypGsc^X}{x^X}\,\cF^{-\Nseg} \,\oLGsc^X\, \eto{\scFX(x^X)}\,,
\end{equation}
where the semicoherent line-odds in detector $X$ is
\mbox{$\oLGsc^X \equiv \probI{\HypLsc^X}/\probI{\HypGsc^X}$}.
Similarly to Eq.~\eqref{eq:pHL_avg}, the posterior probability for the semicoherent
line-hypothesis $\HypLsc$ is obtained as
\begin{equation}
  \label{eq:pHLsc}
  \prob{\HypLsc}{\dVx} = \prob{\HypGsc}{\dVx} \, \cF^{-\Nseg} \,\oLGsc\, \avgX{\rsc^X\,\eto{\scFX(x^X)}}\,,
\end{equation}
where in analogy to Eqs.~\eqref{eq:sumlX} and \eqref{eq:rX} we define
\begin{align}
  \oLGsc &\equiv \frac{\probI{\HypLsc}}{\probI{\HypGsc}} = \sum_X \oLGsc^X\,, \label{eq:oLG_sc}\\
  \rsc^X &\equiv \frac{\oLGsc^X}{\oLGsc/\Ndet}\,.\label{eq:rX_sc}
\end{align}
The posterior probability for the extended noise hypothesis,
\begin{equation}
  \label{eq:4}
  \HypNsc \equiv \left( \HypGsc \OR \HypLsc \right) \,,
\end{equation}
is therefore given by
\begin{equation}
  \label{eq:pHNsc}
  \prob{\HypNsc}{\dVx} = \prob{\HypG}{\dVx} \left( 1 + \cF^{-\Nseg} \,\oLGsc \,\avgX{\rsc^X \eto{\scFX(x^X)}}
\right).
\end{equation}

We can now define a semicoherent line-veto statistic, namely,
\begin{equation}
  \label{eq:OSLsc}
  \OSLsc(\dVx) \equiv \frac{\prob{\HypSsc}{\dVx}}{\prob{\HypLsc}{\dVx}} =
  \oSLsc\, \frac{ \eto{\scF(\dVx)}}{\avgX{\rsc^X\,\eto{\scFX(x^X)} } }\,,
\end{equation}
and a semicoherent line-robust detection statistic as
\begin{equation}
  \label{eq:OSNsc_initial}
  \OSNsc(\dVx) \equiv \frac{\prob{\HypSsc}{\dVx}}{\prob{\HypNsc}{\dVx}} = \left[\OSGsc^{-1}(\dVx) +
  \OSLsc^{-1}(\dVx)\right]^{-1}\!\!.
\end{equation}
The latter can be written explicitly as
\begin{equation}
  \label{eq:OSNsc_final}
  \OSNsc(\dVx) = \oSNsc\,\frac{  \eto{\scF(\dVx) } }
  {(1-\lineprobsc)\,e^{\scFtho} + \lineprobsc \avgX{\rsc^X \eto{\scFX(x^X)} } }\,,
\end{equation}
with (semicoherent) line probability
\begin{equation}
  \label{eq:lineprob_sc}
  \lineprobsc \equiv \frac{\oLGsc}{1 + \oLGsc} = \prob{\HypLsc}{\HypNsc} \;\in\; [0, 1]
\end{equation}
and, in analogy to Eq.~\eqref{eq:Fth0}, a prior cutoff parametrization of
\begin{equation}
  \label{eq:scFtho}
  \scFtho \equiv \ln \cF^{\Nseg}\,.
\end{equation}
Similarly to Eq.~\eqref{eq:OSN_Fstar}, we can therefore write this equivalently as
\begin{equation}
\label{eq:OSNsc_Fstar}
\OSNsc(\dVx) = \oSLsc\,\frac{\eto{\scF(\dVx)}}{\eto{\scFth} + \avgX{\scrX \eto{\scFX(x^X)}}}\,,
\end{equation}
where the semicoherent transition scale $\scFth$ is defined as
\begin{equation}
  \label{eq:scFth}
  \scFth \equiv \scFtho - \ln {\oLGsc}\,,
\end{equation}
by generalizing Eq.~\eqref{eq:Fth}. Hence, we find that $\OSNsc(\dVx)$ transitions from the standard
semicoherent statistic $\OSGsc(\dVx)\propto\eto{\scF}$ to the line-veto statistic $\OSLsc(\dVx)$ when
\begin{equation}
  \label{eq:denomTermsTransition_sc}
  \avgX{\scrX \eto{\scFX}} \sim \eto{\scFth}\,.
\end{equation}

We can rewrite the log-odds as
\begin{equation}
  \begin{split}
  \label{eq:logOSNsc}
  \ln &\OSNsc(\dVx) = \ln \oSLsc + \scF(\dVx) - \scFmaxG(\dVx)\\
  &- \ln \left( \eto{\scFth-\scFmaxG(\dVx)} + \avgX{\rsc^X \eto{\scFX(x^X)-\scFmaxG(\dVx)} } \right)\,,
  \end{split}
\end{equation}
with  $\scFmaxG(\dVx) \equiv \max\left( \scFth,\,\scFX(x^X) + \ln \scrX \right)$.
Note that in the semicoherent case we typically deal with much larger numerical values of $\scF$ [due to its
definition as a sum over segments in Eq.~\eqref{eq:avFstat}]. However, the
logarithmic correction term is still of order unity. This implies that the transition from
$\OSGsc(\dVx)$ to the line-veto odds $\OSLsc(\dVx)$ is expected to be sharper than in the coherent case of
Eq.~\eqref{eq:logOSN}.

Incorporating the ad-hoc $\F$-statistic consistency veto discussed in
Sec.~\ref{sec:line-veto-stat}, we can define a semicoherent \scFvetoName{} as
\begin{equation}
  \label{eq:scFveto}
  \scFveto(\dVx) \equiv \left\{
    \begin{array}{cc}
      \scF(\dVx)\; & \textrm{if } \scF(\dVx) \ge \max_{X}\{\scF^X(x)\}\,,\\
      0        & \textrm{otherwise}\,.
    \end{array}
  \right.
\end{equation}

\section{Choice of priors}
\label{sec:tuning}

The new line-veto and line-robust statistics derived in this paper depend on some prior
parameters which need to be specified. We will now discuss a way to set their values.

The coherent statistics described in Sec.~\ref{sec:line-veto-stats-coh} are simply special cases of the
semicoherent expressions given in Sec.~\ref{sec:semicoherent} for $\Nseg=1$. Hence, in the
following we can use the semicoherent notation without loss of generality.

The pure line-veto statistic $\OSLsc(\dVx)$ of Eq.~\eqref{eq:OSLsc} seems, at first glance, to have $\Ndet$
free parameters.
However, with the sum constraint \eqref{eq:rX} on the line-probability weights $\scrX$ and the fact that
the overall prior odds $\oLGsc$ only enter through the proportionality factor $\oSLsc$, this reduces to an
effective $\Ndet-1$ parameters.
Here we make use again of the fact that all monotonic functions of a test statistic are equivalent in the
Neyman-Pearson sense.

The line-robust statistic $\OSNsc(\dVx)$ depends on the prior odds $\oLGsc$ and on the amplitude-prior cutoff
parameter $\cF$, not just as mere prefactors.
However, these two prior parameters only appear in $\OSNsc$ through the combination
$\scFth \equiv \scFtho - \ln {\oLGsc}$ as defined in Eq.~\eqref{eq:scFth}. Therefore, $\OSNsc$ effectively has
$\Ndet$ free parameters.

While the prior odds $\oLGsc$ have a clear intuitive interpretation, this is not the case for the
prior amplitude cutoff parameter $\cF$ and thus for $\scFtho $, as defined in Eq.~\eqref{eq:scFtho}.
This parameter results from the rather unphysical choice of the amplitude prior in Eq.~\eqref{eq:priorA}, as
discussed in more detail in Refs.~\cite{prix09:_bstat,prix11:_transient}.
Hence, a certain amount of empirical ``tuning'' will be required to determine a reasonable value
for $\scFtho$, which we will discuss in Sec.~\ref{sec:choosing-prior-value}.

\subsection{Proxy estimate of prior line probabilities from the data}
\label{sec:estimate-line-probs}

A maximally uninformative choice for the line-priors would be $\scrX = 1$ and $\oLGsc=1$, where the presence
of lines is considered just as likely as pure Gaussian noise and all detectors are equally likely to be
affected by lines.
A more informed choice should be based on prior characterization of the detectors.

A practical way to achieve this is to judiciously use the observed data $\dVx$ for a simple ``proxy'' estimate
of $\oLGsc^X$.
Empirically we find promising results when adopting the {line-flagging} method of Ref.~\cite{wette09:_thesis}.
We use data from all frequency bins potentially contributing to the detection statistics in a given
search band.
We compute the time-averaged normalized power over these bins and count how many exceed a
predetermined threshold. The measured fraction of such outliers is used as a proxy estimate for the
prior line probability.

More specifically, the data for $\F$-statistic searches is usually prepared in the form of \emph{Short Fourier
Transforms} (SFTs) of the original time-domain data, conventionally spanning
stretches of duration $\Tsft = 1800\,\sec$ (e.g., see Ref.~\cite{krishnan04:_hough}).
We compute the normalized average SFT power $\Psft^X(f)$ for each detector $X$ as (e.g., see
Ref.~\cite{abbott2004:_geoligo})
\begin{equation}
  \label{eq:Psft}
  \Psft^X(f) \equiv \frac{2}{\Nsft\,\Tsft} \sum_{\alpha=1}^{\Nsft}
                    \frac{ \left| \widetilde{x}_\alpha^X(f)\right|^2 }{\SnXal(f)}\,,
\end{equation}
where the sum is over all $\Nsft$ SFTs, $\widetilde{x}_\alpha^X(f)$ and $\SnXal(f)$ denote the
Fourier-transformed data and the noise PSD in the $\alpha$th SFT, respectively.

We estimate the prior line probability $\lineprobsc^X$ for that frequency band as
\begin{equation}
  \label{eq:lineestimator}
  \lineprobsc^X = \frac{\Ncrossed^X}{\Nbins}\,,
\end{equation}
where $\Ncrossed^X$ is the number of bins $\in [0, \Nbins]$ for which $\Psft^X(f)$ crossed the threshold
$\Psftthr^X$. A typical band is of the order of $100\,\mHz$ wide, corresponding to a few hundred bins.
The threshold $\Psftthr^X$ is chosen empirically to be safely above the typical noise fluctuations in the data.

From $\lineprobsc^X$ the prior line odds may be computed as
\begin{equation}
  \label{eq:olg_estimate}
  \oLGsc^X = \frac{\lineprobsc^X}{1 - \lineprobsc^X}\,,
\end{equation}
which also fully specifies $\scrX$ and $\oLGsc$ via Eqs.~\eqref{eq:oLG_sc} and \eqref{eq:rX_sc}.

We determine the threshold $\Psftthr^X$ by fixing a certain false-alarm probability $\pFAl$.
For large $\Nsft$ this can be computed approximately from a Gaussian distribution with unit mean and standard
deviation $\sigma=1/\sqrt{\Nsft}$.

\begin{figure}[h!tbp]
  \includegraphics[width=\columnwidth]{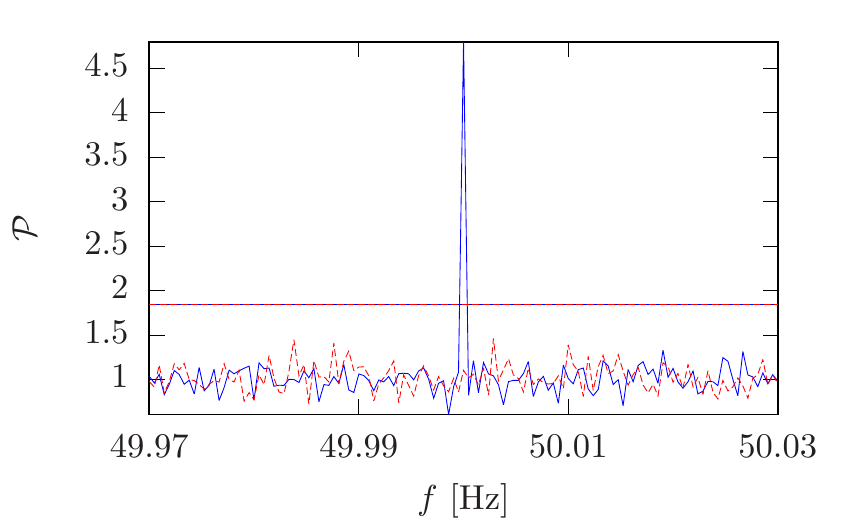}
  \caption{
   \label{fig:tuning_normSFTpower_example}
   Example of the normalized SFT power $\Psft^X(f)$ as a function of frequency $f$ for LIGO $\LHO$
   (solid) and $\LLO$ (dashed) for simulated Gaussian data containing a line in $\LHO$.
   The horizontal line shows the threshold $\Psftthr$ at a false-alarm level of $\pFAl=10^{-9}$.}
\end{figure}

As an illustrative example, Fig.~\ref{fig:tuning_normSFTpower_example} shows $\Psft^X(f)$ for a
$\sim60\,\mHz$ wide band of simulated Gaussian data consisting of 50 SFTs. The data is
generated with a noise PSD of $\Sn^X = 3 \times 10^{-22}\,\Hz^{-1/2}$ in two detectors
$X\in\{\LHO,\LLO\}$, where $\LHO$ and $\LLO$ stand for the LIGO detectors at Hanford and Livingston,
respectively. A monochromatic stationary line of amplitude $h_0 = 2 \times 10^{-23}\,\Hz^{-1/2}$ at
$50\,\Hz$ is injected in $\LHO$ only. More examples from real data are presented in
Sec.~\ref{sec:tests_realdata}.

We stress that the line-flagging procedure proposed here is not meant to yield a direct estimator of
$\lineprobsc^X$ but rather to provide an indication for the presence of lines based on spectral
features that can be robustly identified.

For instance, observing no threshold crossings in the average SFT power $\Psft$ does not necessarily imply
that the $\F$-statistic could not be affected by instrumental artifacts, while seeing many outliers in $\Psft$
does not always yield high values of $\F$.
Hence we will not consider values of $\oLGsc^X$ that suggest more confidence than seems justifiable, and
truncate its range to
\begin{equation}
  \label{eq:olg_trunc}
  \oLGsc^X \in [0.001,\; 1000]\,.
\end{equation}

For the example simulated data set used in Fig.~\ref{fig:tuning_normSFTpower_example} we can detail the
method as follows: there are 127 frequency bins in the band considered, and for the threshold
$\Psftthr^\LHO=\Psftthr^\LLO=\Psftthr(\pFAl=10^{-9},\Nsft=50)\approx1.84$ there is a single crossing in $\LHO$
and none in $\LLO$. Hence, we estimate the line priors as
$\oLGsc^\LHO=\mathrm{max}\left(0.001,\tfrac{1/127}{1-1/127}\right)\approx0.008$ and
$\oLGsc^\LLO=\mathrm{max}\left(0.001,\tfrac{0/127}{1-0/127}\right)=0.001$.

We believe that this data-dependent prior estimation is not prone to the
``sample reuse fallacy'' \cite{jaynes:_logic_of_science}.
The reason is that the proxy estimate for $\oLGsc^X$ is
sufficiently \emph{independent} from the posterior for the line
hypothesis $\HypL$, as they are derived from data sets with effectively very little data in common.
The line hypothesis $\HypL$ (being based on the signal hypothesis $\HypS$) describes a
narrow-band signal, which in each half-hour SFT is confined to a few bins.
In fact the current $\F$-statistic implementation \cite{prix:_cfsv2} uses only $16$ frequency bins per SFT to
construct the detection statistic, and they are very heavily weighted toward a few central ones.
On the other hand, the line-flagging prior estimate uses $\sim\Ord{100-200}$ frequency bins and each counts
equally in the estimate.
Furthermore, the results in Sec.~\ref{sec:tests_realdata} show that this procedure appears to be ``safe'' also
in the presence of (injected) signals.

\subsection{Empirical choice of transition scale \texorpdfstring{$\scFth$}{F*}}
\label{sec:choosing-prior-value}

An additional free parameter in the line-robust statistic $\OSNsc(\dVx)$, as expressed in
Eq.~\eqref{eq:OSNsc_Fstar}, is the transition scale $\scFth= \scFtho - \ln\oLGsc$ of Eq.~\eqref{eq:scFth}.

As discussed in Sec.~\ref{sec:limiting-behavior}, $\scFth$ sets the scale (in terms of $\scFX$) for the
transition of $\OSNsc$ from the signal-versus-Gaussian-noise odds $\OSGsc\propto\eto{\scF}$ (for
$\scFX \ll \scFth$) to the signal-versus-line odds $\OSLsc$ (for $\scFX \gg \scFth$).

Thus we can interpret $\scFtho$ as the transition scale in the case of even prior odds, i.e., $\oLGsc=1$,
between the line and Gaussian-noise hypotheses.

The effect of $\oLGsc$, which we estimate with the method described in the previous section, is to shift the
transition scale up or down from this baseline, depending on whether prior knowledge gives lines lower or
higher odds, respectively.

We can also express $\scFtho$ in terms of a Gaussian-noise false-alarm probability, denoted as
$\pFAtho$:
  \begin{equation}
   \label{eq:scFtho_pFA}
   \pFAtho = \prob{\scF^X > \scFtho}{\HypG}
  \end{equation}
This follows a central $\chi^2$-distribution with $4\Nseg$ degrees of freedom.
We find it useful to fix a value for $\pFAtho$ and use it to determine $\scFtho\left(\pFAtho,\Nseg\right)$.

On the one hand, we want $\scFtho$ to be low enough ($\pFAtho$ high enough) to suppress even weak lines, but
not so low as to compromise the performance in Gaussian noise.
When most of the data is approximately Gaussian (as is typically the case for CW
searches,~\cite{abbott2004:_geoligo,aasi13:_eathS5,behnke2013:_phdthesis}), a reasonable choice is to
use the lowest $\scFtho$ (highest $\pFAtho$) that does not yet adversely affect the detection power in
Gaussian noise.
In practice, we resort to an empirical choice of $\pFAtho$ based on Monte-Carlo simulations on a small subset
of Gaussian or near-Gaussian data.

\section{Performance tests}
\label{sec:tests}

Here we will discuss the detection efficiency of the statistics introduced in
the previous sections for a population of signals embedded in different types of noise.
In order to do this we use two different and somewhat complementary approaches:
(i) fully ``synthetic'' simulations, which allow for efficient large-scale explorations under idealized
conditions, and (ii) injections of simulated signals into LIGO S5 data containing instrumental
artifacts.

We compare the performance of the following statistics (the second equation always refers to the
corresponding semicoherent version):
\begin{enumerate}[(1)] \itemsep1pt \parskip0pt
 \item Standard multidetector $\F$-statistic, Eqs.~\eqref{eq:Fstat} and \eqref{eq:avFstat}
 \item \FvetoName{}, Eqs.~\eqref{eq:Fveto} and \eqref{eq:scFveto}
 \item Line-veto statistic $\OSL$, Eqs.~\eqref{eq:OSL} and \eqref{eq:OSLsc}
 \item Line-robust statistic $\OSN$, Eqs.~\eqref{eq:OSN_Fstar} and \eqref{eq:OSNsc_Fstar}
\end{enumerate}
In the case of the line-robust statistic $\OSN$ we use different transition scales $\Ftho$ corresponding to
false-alarm levels $\pFAtho$, which we denote as
\begin{equation}
  \label{eq:7}
  \OSNpFA{-n}(\dVx) \equiv \OSN(\dVx;\; \pFAtho=10^{-n})\,.
\end{equation}
In the following tests we use $\OSNpFA{-1}$, $\OSNpFA{-3}$, and $\OSNpFA{-6}$, corresponding to
transition-scale false-alarm levels of $\pFAtho=10^{-1},10^{-3},10^{-6}$, respectively.

In order to assess the importance of the choice of prior line odds $\oLG^X$, we consider two cases:
\begin{enumerate}[(i)] \itemsep1pt \parskip0pt
 \item Uninformative priors, i.e., $\oLG^{X} = 1$ for all $X$: the  corresponding ``untuned'' statistics are
       denoted as $\utOSL$ and $\utOSNpFA{-n}$.
 \item Line priors $\oLG^X$ using prior information on the line population: the corresponding ``tuned''
   statistics are denoted as $\OSL$ and $\OSNpFA{-n}$, respectively.
\end{enumerate}

\subsection{Tests using synthetic draws}
\label{sec:tests_simdata}

In this section, for simplicity, we consider only the coherent case (cf.~Sec.~\ref{sec:line-veto-stats-coh}).
Using the synthesizing approach described in Refs.~\cite{prix09:_bstat,prix11:_transient}, one can directly
generate random draws of the various statistics of interest for pure noise and for noise containing a signal.

The synthesizing method consists in generating random draws of the $\{x^X_\mu\}$ of Eq.~\eqref{eq:xmuMmunu}
using their known (multivariate) Gaussian distribution.
From these we compute the $\F$- and $\F^X$-statistics from Eq.~\eqref{eq:Fstat}, $\OSL$ from
Eq.~\eqref{eq:OSL} and $\OSN$ from Eq.~\eqref{eq:OSN_Fstar}.
In the following we refer to each draw of $\{x^X_\mu\}$ together with the resulting statistics as a
\emph{candidate}.

We generate the noise draws in such a way that a fraction $\linefrac$ contains a line according to
$\HypL$ of Eq.~\eqref{eq:hypL}, namely a CW signal in a single detector. The remaining fraction $1-\linefrac$
of noise draws follows the Gaussian-noise hypothesis $\HypG$ of Eq.~\eqref{eq:gaussian}.
In the following we refer to $\linefrac$ as the \emph{line contamination}.

From the noise draws we estimate for each statistic a threshold corresponding to a particular false-alarm
probability $\pFA$. Applying this threshold to the signal candidates yields the detection
probability $\pDet(\pFA)$ for each statistic at the false-alarm level $\pFA$.
This is known as the \emph{receiver operator characteristic} (ROC).

The strength of the injected signals is characterized by the (multidetector) \emph{signal-to-noise ratio}
$\snrS$, defined in the usual way \cite{jks98:_data} as
\begin{equation}
  \label{eq:snrS}
 \snrS^2 \equiv \scalar{h}{h} = \Amp^\mu \M_{\mu\nu} \Amp^\nu \,.
\end{equation}
This is related to the expectation value of the $\F$-statistic as $E[2\F]_{\HypS}=4+\snrS^2$.
As shown in Appendix \ref{sec:expect-f-stat}, for a line according to $\HypL$ in detector $Y$, the
expectation value of the multidetector $\F$-statistic is approximately
\begin{equation}
  \label{eq:exp_F_line}
  \expect{2\F}_{\HypL} \approx 4 + \frac{1}{\Ndet}\,\snrL^2\;\;\text{with}\;\;
  \snrL^2 \equiv \Amp_Y^\mu\M^Y_{\mu\nu}\Amp_Y^\nu\,,
\end{equation}
where we refer to the (single-IFO) SNR $\snrL$ as the ``line SNR''.

\begin{figure}[b!]
  \includegraphics[width=1.05\columnwidth,clip]{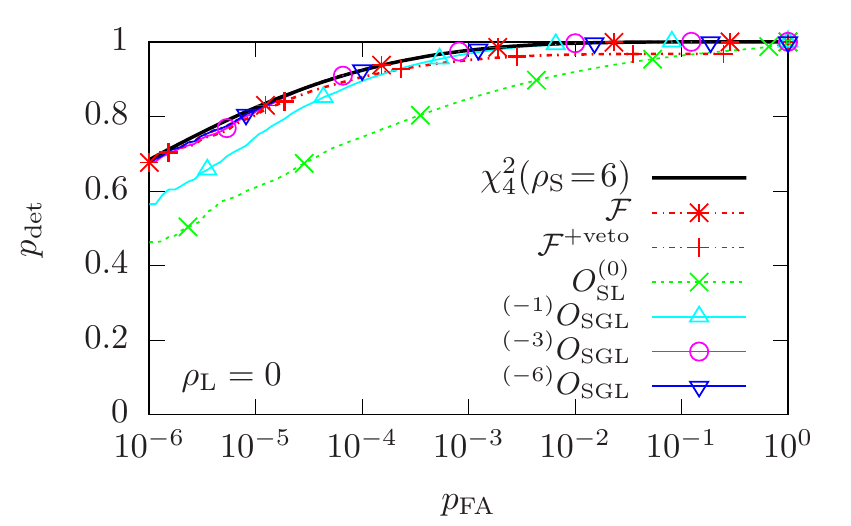}
  \caption{
    \label{fig:newSynth_Gauss}
    Detection probability $\pDet$ as a function of false-alarm $\pFA$ of different synthesized statistics,
    for a signal population of fixed SNR of $\snrS=6$ in pure Gaussian noise ($\linefrac=0$, $\snrL=0$).
    Statistical errors are similar to the line width.
  }
\end{figure}

The signal candidates are generated for a fixed SNR of $\snrS=6$, and a data length of $T=25\,\hours$ is
assumed.
This signal strength is chosen to be representative of reasonably detectable signals in a wide-parameter-space
search.
In such a search we would require a low (single-trial) false-alarm threshold $\pFA$ in order to consider a
candidate as significant.
The choice of $\snrS=6$ corresponds to a detection probability of $\pDet\approx70\%$ at a false-alarm
probability of $\pFA=10^{-6}$ in Gaussian noise (for example, see Fig.~\ref{fig:newSynth_Gauss}).

The signal amplitude parameters are drawn uniformly in $\cosi\in[-1,1]$, $\psi\in[-\pi/4,\pi/4]$ and
$\phio\in[0,2\pi]$.
The sky position is drawn isotropically over the sky, and $(h_0/\sqrt{\Sn})$ is determined by the fixed signal
SNR of $\snrS=6$ according to Eq.~\eqref{eq:snrS}.
The line draws use the same prior distributions, but the signal is added to only one detector, and
$(h_0/\sqrt{\Sn})$ is determined by fixing a (single-IFO) line SNR $\snrL$ according to
Eq.~\eqref{eq:exp_F_line}.

In each simulation we generate $10^7$ noise candidates and $10^7$ noise+signal candidates for two detectors,
LIGO $\LHO$ and $\LLO$. These detectors are assumed here to have identical sensitivity.
Lines are only injected into $\LHO$ without loss of generality.
We consider three examples of noise populations:
\begin{enumerate}[(i)] \itemsep1pt \parskip0pt
  \item pure Gaussian noise without lines ($\linefrac=0$, $\snrL=0$)
  \item 10\% line contamination in H1 ($\linefrac^{\LHO}=0.1,\;\linefrac^{\LLO}=0$) with line SNR of $\snrL=9$,
  \item 10\% line contamination in H1  ($\linefrac^{\LHO}=0.1,\;\linefrac^{\LLO}=0$) with line SNR of $\snrL=15$.
\end{enumerate}
The line SNR of $\snrL=9$ corresponds to lines that are marginally stronger than the injected signals, namely,
$\expect{2\F}_{\HypL}\approx 44.5$ from Eq.~\eqref{eq:exp_F_line}), while $\expect{2\F}_{\HypS} = 40$ from
Eq.~\eqref{eq:14}).
The lines with $\snrL=15$ are substantially stronger (namely $\expect{2\F}_{\HypL} \approx 117$) than the
injected signals.

Note that for the synthesized statistics we cannot use the line-prior estimation method for $\oLG^X$ of
Sec.~\ref{sec:estimate-line-probs}. Instead we assume ``perfect tuning'': in
the Gaussian-noise example we set $\oLG^X=0.001$ for $X=\LHO,\LLO$, and in the two line examples we use
$\lineprob^{\LHO}=\linefrac^{\LHO}=0.1$ (therefore $\oLG^\LHO=1/9$) and $\oLG^\LLO=0.001$ (no lines were
injected into L1).

\begin{figure}[b]
    \raggedright (a)\\\vspace*{-0.5cm}
    \includegraphics[width=\columnwidth,clip]{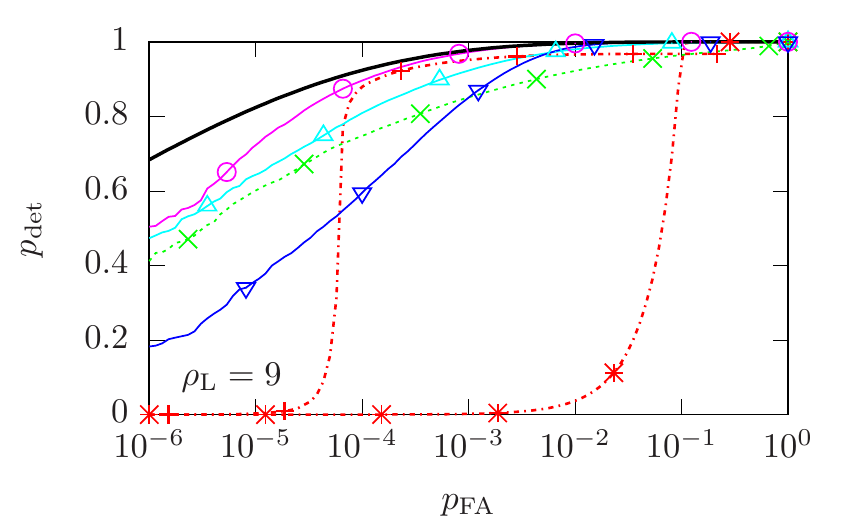}
    \raggedright(b)\\\vspace*{-0.5cm}
    \includegraphics[width=\columnwidth]{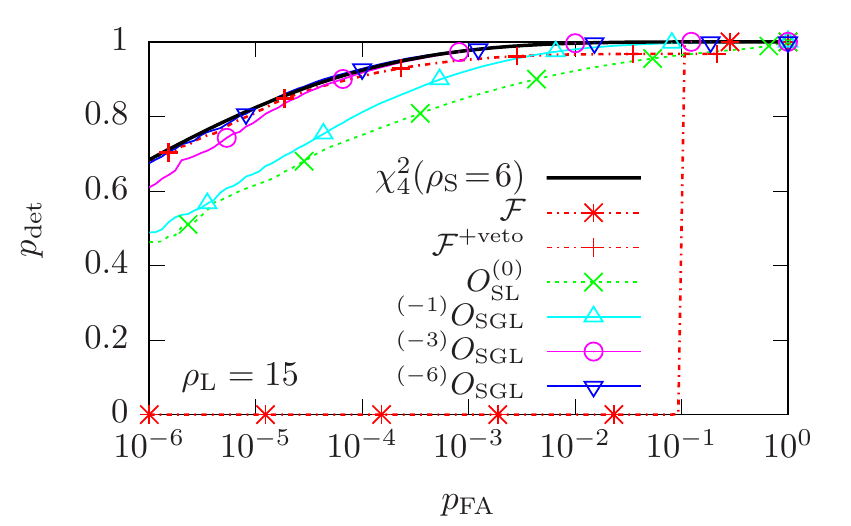}
  \caption{
    \label{fig:newSynth_Lines}
    Detection probability $\pDet$ as a function of false-alarm $\pFA$ for different synthesized statistics,
    for a signal population with fixed SNR of $\snrS=6$ in Gaussian noise with 10\% line contamination,
    with line-SNR of (a) $\snrL=9$ and (b) $\snrL=15$.
    Statistical errors are similar to the line width.
  }
\end{figure}

In Gaussian noise the coherent $\F$-statistic is close to optimal \cite{jks98:_data,prix09:_bstat}, and follows
a $\chi^2$ distribution with 4 degrees of freedom and noncentrality parameter $\snrS^2$, which we denote as
$\chi^2_4(\snrS)$. This is plotted as a thick solid line in Figs.~\ref{fig:newSynth_Gauss} and
\ref{fig:newSynth_Lines} for the signal population of $\snrS=6$.

In the Gaussian-noise example shown in Fig.~\ref{fig:newSynth_Gauss}, the $\F$-statistic follows closely
the theoretical prediction, while the (untuned) line-veto statistic $\utOSL$ is notably less powerful.
The line-robust statistics $\OSNpFA{-n}$ increasingly approach the $\F$-statistic performance with decreasing
$\pFAtho$, i.e., increasing transition scale $\Ftho$.
In particular, starting from $\OSNpFA{-3}$ (corresponding to a transition scale of $\Ftho\approx 9.23$),
there are no appreciable losses in detection probability $\pDet$ over the false-alarm range $\pFA\in
[10^{-6},1]$.

At low $\pFA$, the \FvetoName{} performs almost optimally, while
there are some losses above $\pFA\gtrsim10^{-4}$. These are due to $\Fveto$ containing intrinsic upper
bounds on the achievable $\pFA$ and $\pDet$ as a result of vetoing a finite fraction of candidates.
For a practical GW analysis, where low $\pFA$ are required, this behavior is not particularly
relevant.

The performance in the two examples with $10\%$ line contamination is shown in
Fig.~\ref{fig:newSynth_Lines}.
Here the $\F$-statistic is found to perform substantially worse than in Gaussian noise at false-alarm
probabilities below $\pFA\lesssim 0.1$. This is due to the fact that in $10\%$ of the noise cases the
false-alarm threshold is set by the line population, which is either difficult (for $\snrL=9$, left plot) or
almost impossible (for $\snrL=15$, right plot) for the $\F$-statistic to cross for signals with SNR of
$\snrS=6$.

\begin{figure}[b]
    \raggedright (a)\\\vspace*{-0.5cm}
    \includegraphics[width=\columnwidth,clip]{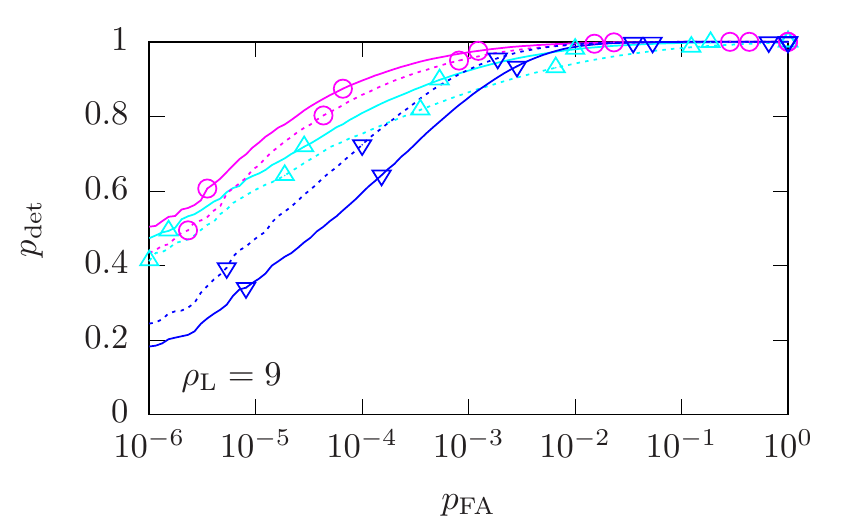}
    \raggedright(b)\\\vspace*{-0.5cm}
    \includegraphics[width=\columnwidth]{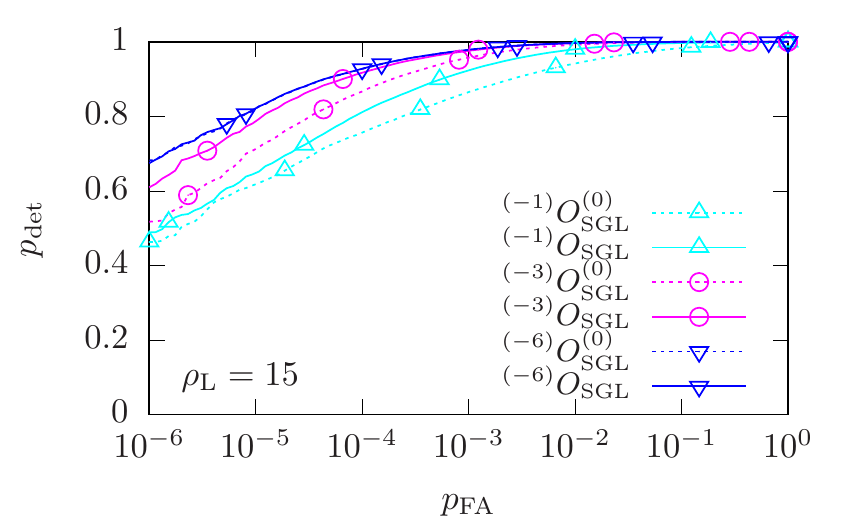}
  \caption{
    \label{fig:newSynth_LinesTuning}
    Comparison of ``tuned'' statistics $\OSNpFA{-n}$ (solid lines) using
    ``perfect knowledge'' line-priors $\{\oLG^\LHO=1/9,\,\oLG^\LLO=10^{-3}\}$ versus ``untuned'' statistic
    $\utOSNpFA{-n}$ (dashed lines) using uninformative line-priors $\oLG^X=1$.
    Detection probability $\pDet$ as a function of false-alarm $\pFA$ of different synthesized statistics,
    for a signal population with fixed SNR of $\snrS=6$ in Gaussian noise with 10\% line contamination,
    with line-SNR of (a) $\snrL=9$ and (b) $\snrL=15$.
    Statistical errors are similar to the line width.
  }
\end{figure}

We observe that the \FvetoName{} starts to fail below false-alarm levels of
$\pFA\lesssim 10^{-4}$ in the case of weaker lines with $\snrL=9$ (see Fig.~\ref{fig:newSynth_Lines}(a)).
This can be understood as follows:
For the $\snrL=9$ line population, we find that a fraction of $\sim6\times10^{-4}$ of line candidates survive
the veto.
Given that lines are present in 10\% of the noise cases, this means that a fraction of
$\sim6\times10^{-5}$ of total noise candidates are line candidates surviving the consistency veto.
Given that these have high $\F$-statistic values, signal candidates can hardly surpass them, and thus the
detection probability drops toward zero at false-alarm probabilities below $\sim6\times10^{-5}$.

The same effect is also present for stronger lines, but the corresponding ``failure'' threshold is
pushed to lower values.
For example, for $\snrL=12$ it would happen only below $\pFA\lesssim10^{-6}$, while for $\snrL=15$ it
is too low to be resolvable by $10^7$ random draws.

The behavior of the line-robust statistics $\OSNpFA{-n}$ depends on the choice of transition scale.
In the case of lines with $\snrL=9$, shown in Fig.~\ref{fig:newSynth_Lines}(a), the statistic $\OSNpFA{-3}$
performs best, while using either lower or higher values of $\pFAtho$ is less powerful at low false-alarm
probabilities.
In the case of stronger lines with $\snrL=15$, shown in Fig.~\ref{fig:newSynth_Lines}(b), the statistic
$\OSNpFA{-6}$ performs almost optimally, with $\OSNpFA{-3}$ performing only slightly worse.

The line-veto statistic $\utOSL$ performs somewhat poorly in all three examples
shown (Figs.~\ref{fig:newSynth_Gauss} and \ref{fig:newSynth_Lines}). This is not surprising, given that at
most $10\%$ of noise draws contain a line, while $\OSL$ would only be optimal for a noise population
consisting exclusively of lines.

Figure~\ref{fig:newSynth_LinesTuning} shows the effect of ``tuning'' the prior line odds
$\oLG^X$, using the same line populations as in Fig.~\ref{fig:newSynth_Lines}. We see that the untuned
statistics $\utOSL$ and $\utOSNpFA{-n}$ using uninformative line-odds $\oLG^X=1$ perform reasonably well
compared to $\OSL$ and $\OSNpFA{-n}$, which are based on ``perfect-knowledge'' tuning.
Note that tuning of $\oLG^X$ can sometimes also {decrease} the detection power of a statistic,
particularly in cases where the choice of the transition scale $\Ftho(\pFAtho)$ is a poor fit to the actual
line population.
This can be seen in the case of $\OSNpFA{-6}$ with lines of SNR $\snrL=9$, as shown in
Fig.~\ref{fig:newSynth_LinesTuning}(a).
In cases where $\Ftho(\pFAtho)$ is a good match to the line population, the tuning of $\oLG^X$ can yield gains
in detection power of up to 5--10\%.

\subsection{Tests using LIGO S5 data}
\label{sec:tests_realdata}

Here we conduct a study using LIGO S5 data sets as noise.
We inject signals and search for them using methods similar to those of actual CW searches.
Instead of ROC curves, i.e., $\pDet(\pFA)$, we present results in terms of the detection efficiency as
a function of signal strength scaled by the total multidetector noise PSD $\Sn$, i.e.,
$\pDet(h_0/\sqrt{\Sn})$.
This form is more suitable to assess improvements in sensitivity, which is typically expressed as the
weakest signal $h_0$ detectable with a certain confidence $\pDet$.
To compute an astrophysically motivated detection probability, these results could in principle be
convolved with an astrophysical prior on $h_0$, if available.

The injection and detection procedure used here is modeled after those commonly employed for estimating upper
limits on $h_0$ in CW searches such as Refs.~\cite{aasi13:_eathS5, aasi2013:_gc-search}.

Simulated CW signals are added to the data using the \texttt{Makefakedata\_v4} code \cite{lalsuite}.
The resulting data set is analyzed both coherently and semicoherently using
\texttt{HierarchSearchGCT} \cite{lalsuite}, a StackSlide implementation based on
the ``global correlations'' method of Ref.~\cite{pletsch2009:_gct}. We have extended this code to also compute
the new statistics $\OSLsc(\dVx)$ and $\OSNsc(\dVx)$, in addition to $\scF(\dVx)$.
For the coherent search we use shorter subsets of the data, and the coherent statistics are simply obtained as
the special case $\Nseg=1$.

The tuning of the line priors $\oLGsc^X$ in $\OSLsc$ and $\OSNscpFA{-n}$ is based on the method described in
Sec.~\ref{sec:estimate-line-probs}, namely, Eqs.~\eqref{eq:lineestimator},~\eqref{eq:olg_estimate}, and
\eqref{eq:olg_trunc}.
As explained in Sec.~\ref{sec:choosing-prior-value}, we fix the transition scale $\scFtho$ of $\OSNsc$
according to its performance in Gaussian noise.
Specifically, we perform injections on simulated Gaussian noise and analyze them as
described below for several values of $\pFAtho$.
We then chose the highest $\pFAtho$ value such that the achieved performance is indistinguishable within
statistical uncertainties from that of the $\F$-statistic.
As a result of this we select $\OSNpFA{-6}$,
with the false-alarm level of $\pFAtho=10^{-6}$ corresponding to a transition scale of
\mbox{$\Ftho(\Nseg=1)\approx16.7$} and \mbox{$\scFtho(\Nseg=84)\approx237.0$}, respectively.

\subsubsection{Data selection}
\label{sec:data-selection}
\begin{figure}[h!tbp]
    \raggedright (\acoh)\\\vspace*{-1cm}
    \includegraphics[width=\columnwidth]{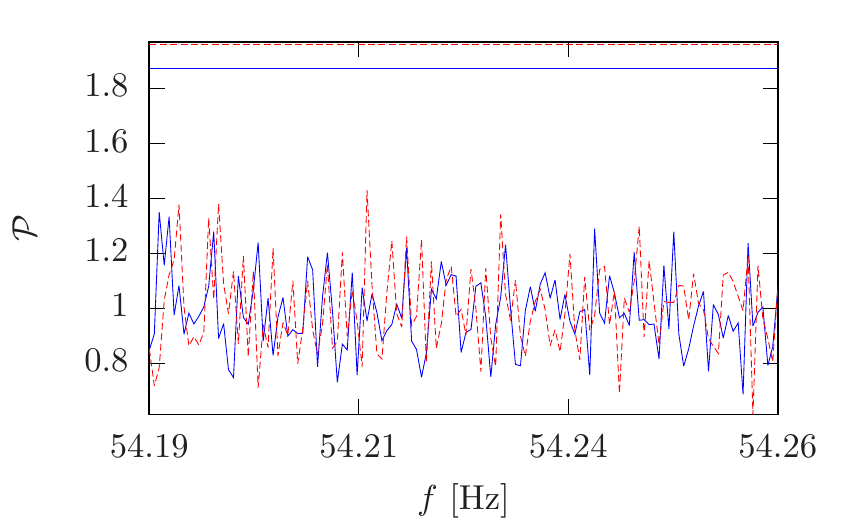} \\
    \raggedright (\bcoh)\\\vspace*{-0.5cm}
    \includegraphics[width=\columnwidth]{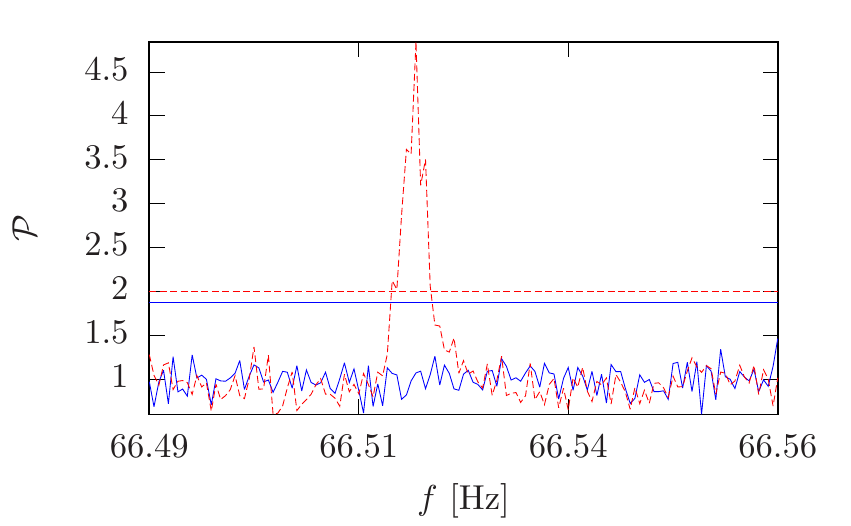}\\
    \raggedright (\ccoh)\\\vspace*{-0.5cm}
    \includegraphics[width=\columnwidth,clip]{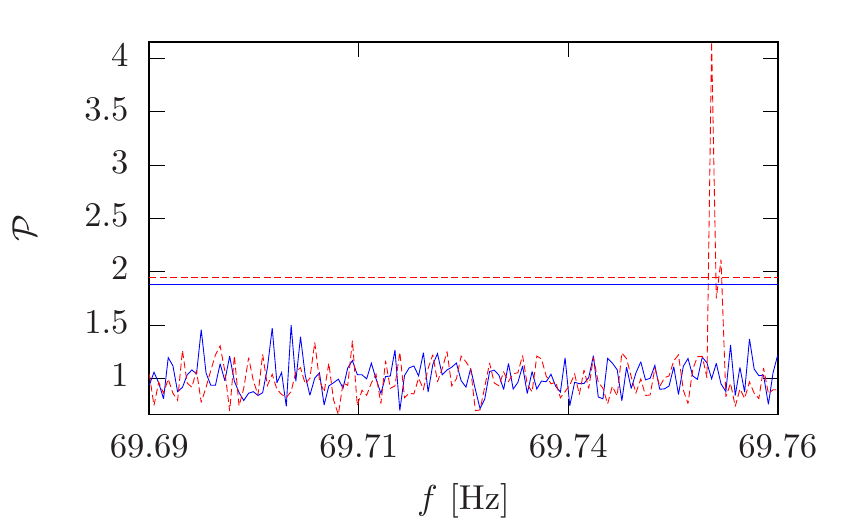} \\
    \raggedright (\dcoh)\\\vspace*{-0.5cm}
    \includegraphics[width=\columnwidth]{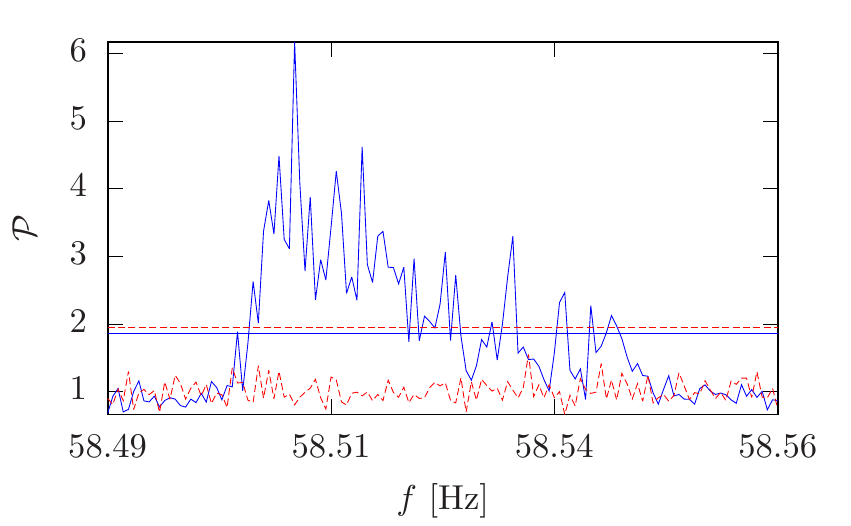}
  \caption{
    \label{fig:tests_realdata_normSFT_coh}
    Normalized average SFT power $\Psft^X(f)$ of Eq.~\eqref{eq:Psft} as a function of frequency $\Freq$
    for LIGO H1 (solid) and L1 (dashed) data used in the coherent searches.
    The horizontal lines mark, for each detector, the threshold $\Psftthr^X$ at $\pFAl=10^{-9}$ used in the
    line prior estimation.
    The panels show:
    (\acoh) a quiet band,
    (\bcoh), (\ccoh) two bands with lines,
    (\dcoh) a band with multiple disturbances. See Table~\ref{tbl:PsftthrCoh} for more details on these data
     sets.}
\end{figure}

We use \numexamplebands\ narrow frequency bands of LIGO S5 data.
These bands are chosen depending on how severely they appear to be affected by lines:
\begin{enumerate}[(a)]
\item a ``quiet'' band where the distribution of the data is very close to Gaussian,
\item a band with a single line in $\LLO$,
\item a band with a single line in $\LLO$, narrower than in (b),
\item a band with multiple disturbances in $\LHO$.
\end{enumerate}
The normalized SFT power $\Psft^X(f)$ of Eq.~\eqref{eq:Psft} for each of the \numexamplebands\ bands is shown
in Fig.~\ref{fig:tests_realdata_normSFT_coh} for the coherent case, and in
Fig.~\ref{fig:tests_realdata_normSFT_semicoh} for the semicoherent case.
More details about these sample frequency bands are given in Tables~\ref{tbl:PsftthrCoh} and
\ref{tbl:PsftthrSC}, respectively.

The data sets are taken from the first year of the LIGO S5 science run.
For the semicoherent searches we use $\Nseg=84$ data segments, spanning $T=25\,\hours$ each, while the
coherent searches use only a single segment.
These segments were originally selected for the Einstein@Home~\cite{EatH} search described in
Ref.~\cite{aasi13:_eathS5}.
Since CW searches on this data have not found any signals
\cite{abbott09:_earlyS5,abadie12:_powerflux,aasi13:_eathS5}, we consider it as a \emph{pure noise} set
for the purpose of this study.

\begin{table*}[h!tbp]
 \begin{tabular}{c c c c c c c c c c c}
 \hline\hline
 Label & $\finj\,[\Hz]$  & $\fsft\,[\Hz]$  & $\tstart$ $[\sec]$ & $\sqrt{\Sn}\,[\Hz^{-1/2}]$ & $\maxnoise{2\F}$ & Detector $X$ & $\Nsft^X$ & $\sqrt{\SnX}\,[\Hz^{-1/2}]$ & $\Psftthr^X$ & $\oLG^{X}$ \\\hline

 (\acoh) & [54.20,\,54.25] & [54.19,\,54.26] & 835120582 & $1.80 \times 10^{-22}$ & 38.04 & H1 & 47 & $2.51 \times 10^{-22}$ & 1.875 & 0.001 \\
 & & & & & & L1 & 39 & $1.48 \times 10^{-22}$ & 1.960 & 0.001 \\[0.1cm]

 (\bcoh) & [66.50,\,66.55] & [66.49,\,66.56] & 844876223 & $1.28 \times 10^{-22}$ & 81.68 & H1 & 47 & $1.19 \times 10^{-22}$ & 1.875 & 0.001 \\
 & & & & & & L1 & 36 & $1.41 \times 10^{-22}$ & 2.000 & 0.073 \\[0.1cm]

 (\ccoh) & [69.70,\,69.75] & [69.69,\,69.76] & 821912087 & $1.42 \times 10^{-22}$ & 122.54 & H1 & 46 & $1.35 \times 10^{-22}$ & 1.884 & 0.001 \\
 & & & & & & L1 & 40 & $1.50 \times 10^{-22}$ & 1.948 & 0.015 \\[0.1cm]

 (\dcoh) & [58.50,\,58.55] & [58.49,\,58.56] & 827366996 & $2.48 \times 10^{-22}$ & 104.46 & H1 & 48 & $2.58 \times 10^{-22}$ & 1.866 & 0.585 \\
 & & & & & & L1 & 40 & $2.40 \times 10^{-22}$ & 1.948 & 0.001 \\[0.1cm]

 \hline\hline
\end{tabular}
 \caption{
   \label{tbl:PsftthrCoh}
   Data used for tests of the coherent statistics in Sec.~\ref{sec:tests_realdata}.
   All data is taken from the first year of the LIGO S5 run.
   CW signals are injected with frequencies $f\in \finj$, while $\fsft$ denotes the SFT frequency
   range used for the search and the prior line estimation.
   Each data set starts at a GPS time of $\tstart$ and spans 25 hours, containing $\Nsft^X$
   SFTs of duration $\Tsft=1800\,\sec$ from each detector.
   The multidetector noise PSD $\Sn$ was obtained as the harmonic mean over SFTs and arithmetic
   mean over frequency bins.
   The column labeled $\maxnoise{2\F}$ shows the corresponding highest multidetector $2\F$ value without
   injections.
   The noise PSD per detector is $\SnX$.
   The column $\Psftthr^X$ gives the threshold on the normalized SFT power $\Psft^X$ at $\pFAl=10^{-9}$,
   which is used to estimate the prior line-odds $\oLG^X$ as described in Sec.~\ref{sec:estimate-line-probs}.
 }
\end{table*}


\begin{table*}[h!tbp]
 \begin{tabular}{c c c c c c c c c c}
 \hline\hline
 Label  & $\finj\,[\Hz]$  & $\fsft\,[\Hz]$ & $\sqrt{\Sn}\,[\Hz^{-1/2}]$ & $\maxnoise{2\avF}$ & Detector $X$ & $\Nsft^X$ & $\sqrt{\SnX}\,[\Hz^{-1/2}]$ & $\Psftthr^X$ & $\oLGsc^{X}$ \\\hline

 (\asc) & [54.20,\,54.25] & [54.15,\,54.30] & $2.09 \times 10^{-22}$ & 6.51 & H1 & 3781 & $2.54 \times 10^{-22}$ & 1.098 & 0.001 \\
 & & & & & L1 & 3456 & $1.81 \times 10^{-22}$ & 1.102 & 0.001 \\[0.1cm]

 (\bsc) & [66.50,\,66.55] & [66.44,\,66.61] & $1.14 \times 10^{-22}$ & 10.83 & H1 & 3781 & $1.35 \times 10^{-22}$ & 1.098 & 0.001 \\
 & & & & & L1 & 3456 & $1.00 \times 10^{-22}$ & 1.102 & 0.047 \\[0.1cm]

 (\csc) & [69.70,\,69.75] & [69.64,\,69.81] & $1.01 \times 10^{-22}$ & 83.48 & H1 & 3781 & $1.15 \times 10^{-22}$ & 1.098 & 0.001 \\
 & & & & & L1 & 3456 & $9.08 \times 10^{-23}$ & 1.102 & 0.017 \\[0.1cm]

 (\dsc) & [58.50,\,58.55] & [58.45,\,58.60] & $2.12 \times 10^{-22}$ & 8.35 & H1 & 3781 & $2.20 \times 10^{-22}$ & 1.098 & 1.743 \\
 & & & & & L1 & 3456 & $2.05 \times 10^{-22}$ & 1.102 & 0.001 \\[0.1cm]

 \hline\hline
\end{tabular}
 \caption{
   \label{tbl:PsftthrSC}
   Data used for tests of the semicoherent statistics in Sec.~\ref{sec:tests_realdata}.
   All data is taken from the first year of the LIGO S5 run, corresponding to the segment
   selection used in an Einstein@Home search (S5R3) \cite{aasi13:_eathS5}, spanning 381.04~days
   starting from GPS epoch $\tstart=818845553$, containing $\Nseg=84$ segments, each 25 hours long.
   The column labeled $\maxnoise{2\avF}$ refers to the highest average multidetector
   $2\avF$ value without injections (the average is over segments).
   The remaining labels are identical to those in Table~\ref{tbl:PsftthrCoh}.
   }
\end{table*}

\subsubsection{Signal injection and detection criterion}
\label{sec:sign-inject-detect}

The search setup used here is different from that of Ref.~\cite{aasi13:_eathS5}, and employs the
\texttt{HierarchSearchGCT} code instead of the Hough-transform~\cite{krishnan04:_hough}.
This code is used in recent and ongoing wide-parameter-space searches such as
Refs.~\cite{aasi2013:_gc-search,EatH}.

The grid spacings in frequency and spin-down are $\delta f \approx 1.6 \times 10^{-6}\,\Hz$ and $\delta\dot{f}
\approx 5.8 \times 10^{-11}\,\Hz/s$, respectively.
The angular sky-grid spacings are approximately $0.15\,\mathrm{rad}$ at $\Freq = 54\,\Hz$, and scale with
frequency as $1/\Freq$.

We find that this template bank yields an average relative loss of SNR$^2$ (also known as mismatch) of $m\sim
0.6$ in the
semicoherent searches and of $m\lesssim 0.05$ in the coherent searches.


We first perform searches on the data without any injections,
covering the whole sky in each of the \numexamplebands{} frequency bands of width $\Delta f = 50\,\mHz$ (see
$\finj$ in Tables \ref{tbl:PsftthrCoh} and \ref{tbl:PsftthrSC}), and a fixed band $[-\Delta\dot{f},\,0]$ in
spin-down $\fdot$, with $\Delta \dot{f} \approx 2.6 \times 10^{-9}\,\Hz/\sec$.

For each of the four statistics $\{\scF,\scFveto,\OSLsc,\OSNsc\}$ we record the loudest noise candidate over
the whole template grid.
A signal will be considered as detected with a given statistic if its highest value exceeds this noise value.
This definition of detection is equivalent to the common method of setting loudest-event upper
limits, employed for example in Ref.~\cite{aasi13:_eathS5}.

The signals are injected using the \texttt{Makefakedata\_v4} code, with signal parameters randomly drawn
from uniform distributions in the sky coordinates $\{\alpha,\delta\}$,
inclination $\cosi$ and polarization angle $\psi$, and at varying signal amplitude $h_0$.
The signal frequency and spin-down are drawn uniformly from the bands used in the noise search.
For each value of $h_0$ we perform 1000 injections.
For each injection we search a small parameter-space volume containing the signal.
This search region consists of a frequency band of $\Delta f = 1\,\mHz$, a spin-down band of
$\Delta \dot{f} \approx
2.3 \times 10^{-10}\,\Hz/\sec$ and the 10 sky-grid points closest (in the metric sense \cite{prix06:_searc})
to the injection.

Note that in (b) and (c) some of these injection searches do not use any data containing the narrow
disturbances. Hence, the statements in this section apply to \emph{bands that contain disturbances}, and not
only to \emph{sets of disturbed candidates}.

\begin{figure}[h!tbp]
    \raggedright (\acoh)\\\vspace*{-1cm}
    \includegraphics[width=\columnwidth]{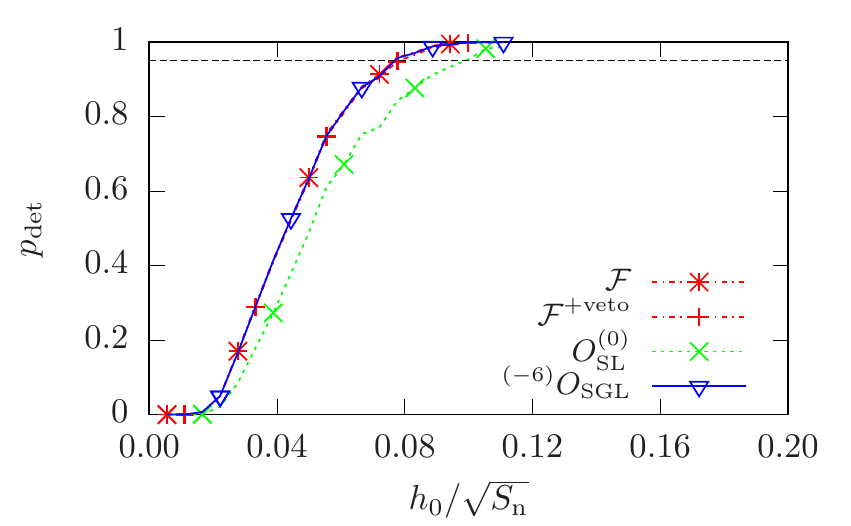} \\
    \raggedright (\bcoh)\\\vspace*{-0.5cm}
    \includegraphics[width=\columnwidth,clip]{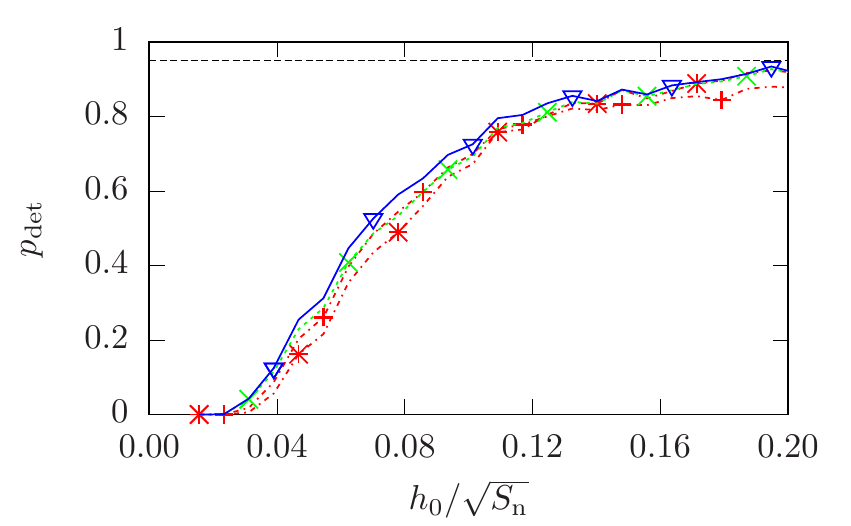} \\
    \raggedright (\ccoh)\\\vspace*{-0.5cm}
    \includegraphics[width=\columnwidth]{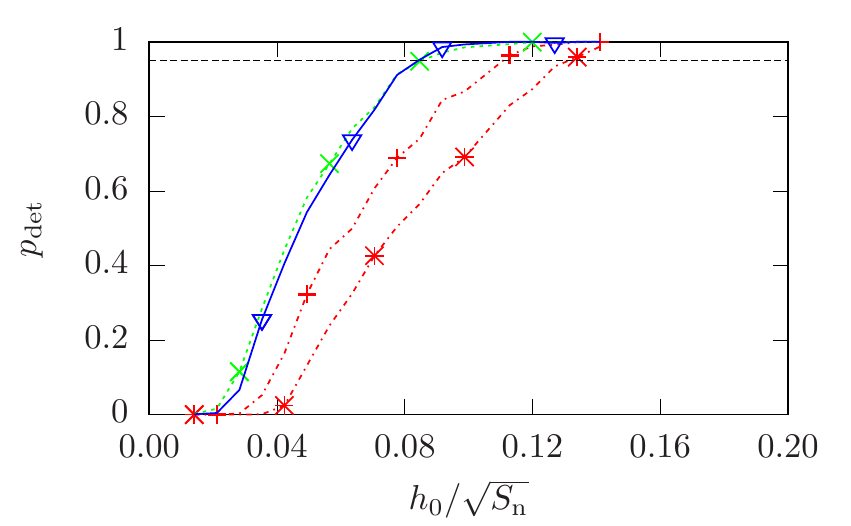}
    \raggedright (\dcoh)\\\vspace*{-0.5cm}
    \includegraphics[width=\columnwidth]{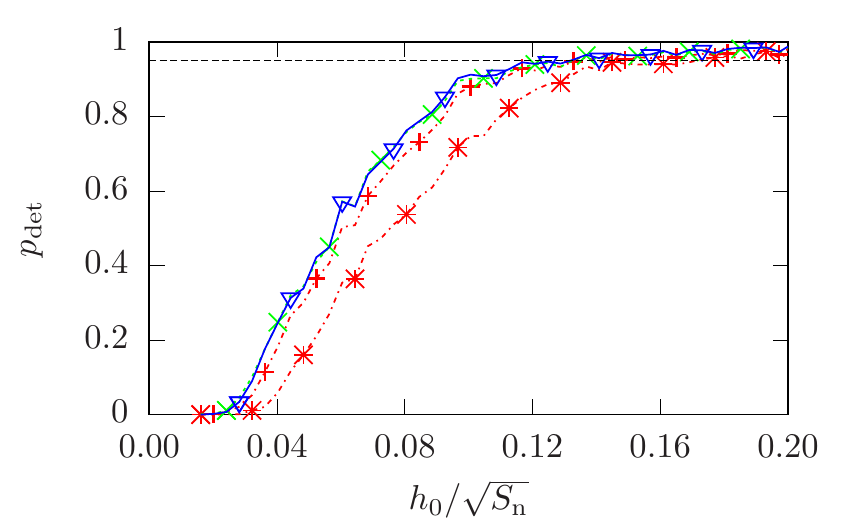}
  \caption{
    \label{fig:tests_realdata_detprobs_coherent}
    Detection efficiency $\pDet$ as a function of scaled signal amplitude $h_0/\sqrt{\Sn}$ for four
    different coherent statistics:
    $\F$, 
    $\Fveto$, 
    $\utOSL$, 
    and $\OSNpFA{-6}$. 
    Statistical errors are similar to the size of the symbols.
    The dashed horizontal line marks the $95\%$ detection probability level.
    The panels show:
    (\acoh) a quiet band,
    (\bcoh), (\ccoh) two bands with lines,
    (\dcoh) a band with multiple disturbances. See Fig.~\ref{fig:tests_realdata_normSFT_coh} and
    Table~\ref{tbl:PsftthrCoh} for more details on these data sets.
  }
\end{figure}


\begin{figure}[h!tbp]
    \raggedright(\asc)\\\vspace*{-1cm}
    \includegraphics[width=\columnwidth]{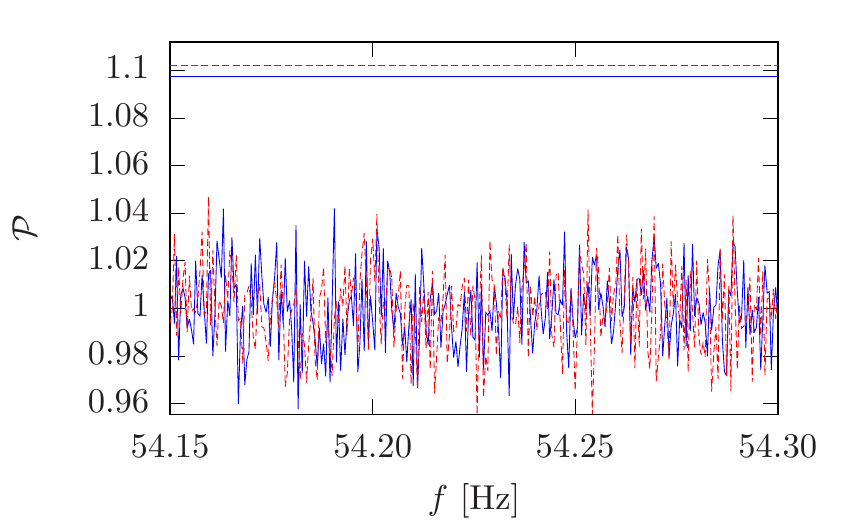} \\
    \raggedright (\bsc)\\\vspace*{-0.5cm}
    \includegraphics[width=\columnwidth]{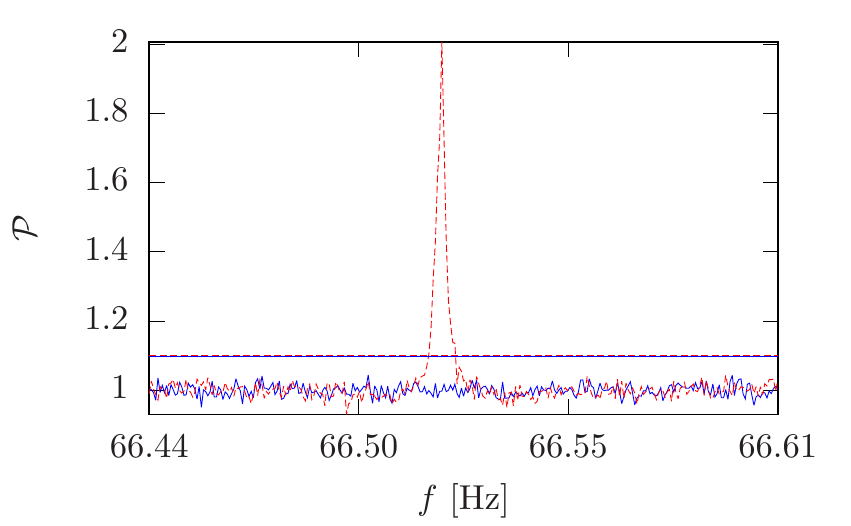} \\
    \raggedright (\csc)\\\vspace*{-0.5cm}
    \includegraphics[width=\columnwidth,clip]{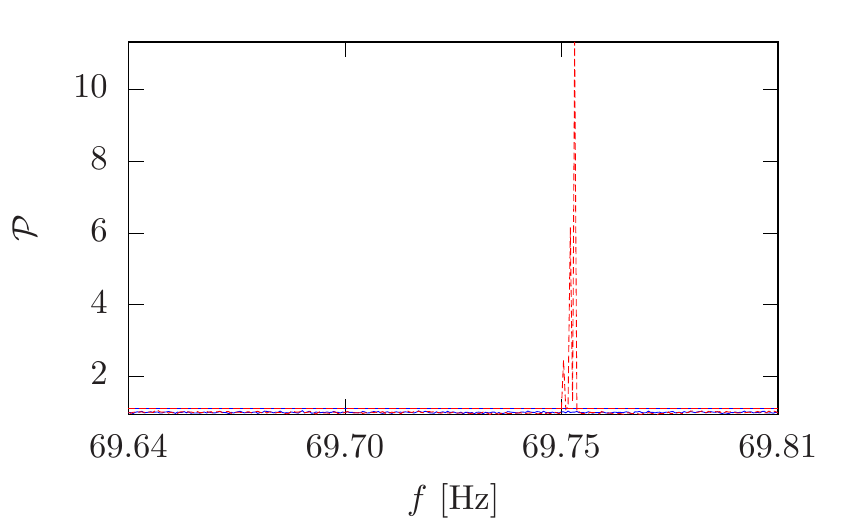}
    \raggedright (\dsc)\\\vspace*{-0.5cm}
    \includegraphics[width=\columnwidth]{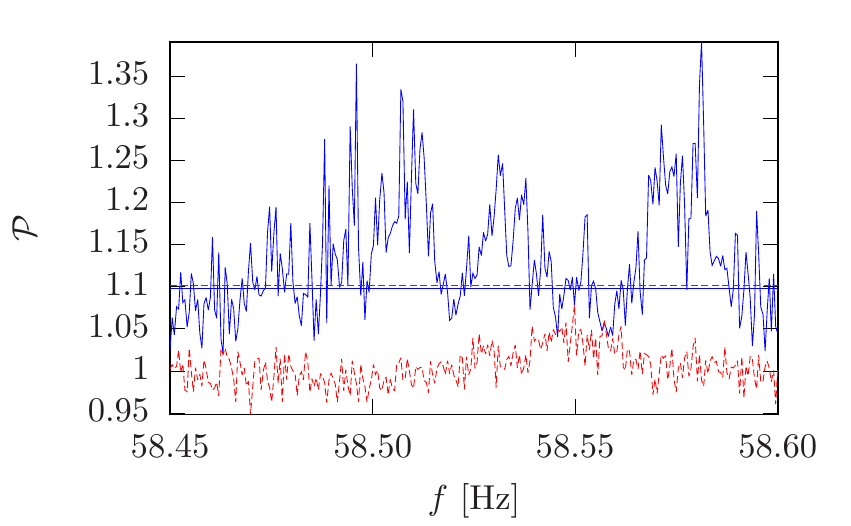}
  \caption{
    \label{fig:tests_realdata_normSFT_semicoh}
    Normalized average SFT power $\Psft^X(f)$ of Eq.~\eqref{eq:Psft} as a function of frequency $\Freq$
    for LIGO H1 (solid) and L1 (dashed) data used in the semicoherent searches.
    The horizontal lines mark, for each detector, the threshold $\Psftthr^X$ at $\pFAl=10^{-9}$ used in the
    line prior estimation.
    The panels show:
    (\asc) a quiet band,
    (\bsc), (\csc) two bands with lines,
    (\dsc) a band with multiple disturbances.
    See Table~\ref{tbl:PsftthrSC} for more details on these data sets.
  }
\end{figure}
\begin{figure}[h!tbp]
    \raggedright (\asc)\\\vspace*{-1cm}
    \includegraphics[width=\columnwidth,clip]{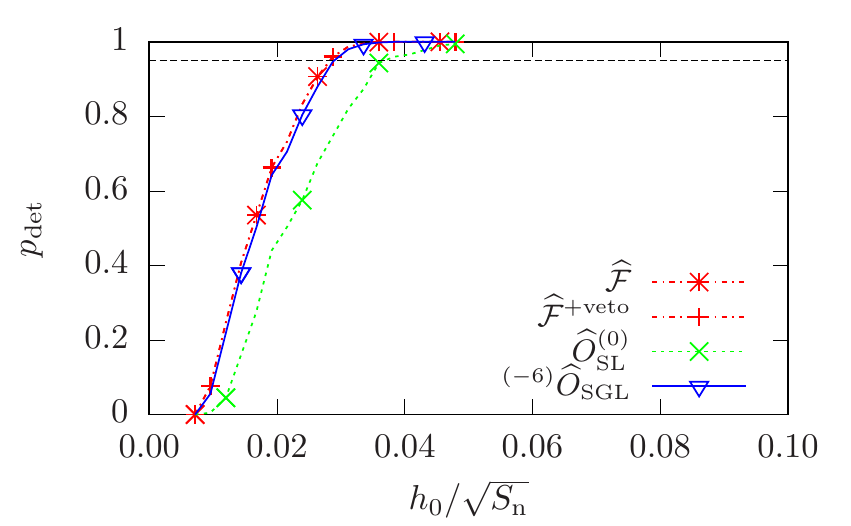} \\
    \raggedright (\bsc)\\\vspace*{-0.5cm}
    \includegraphics[width=\columnwidth,clip]{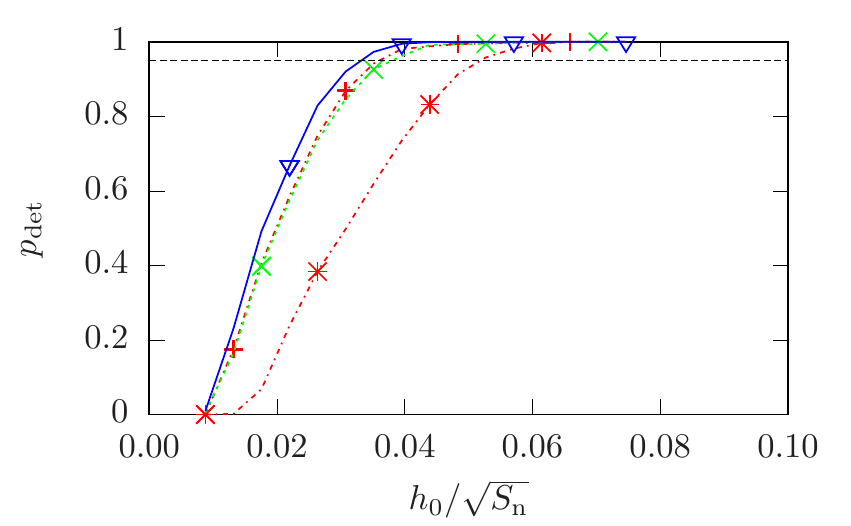} \\
    \raggedright (\csc)\\\vspace*{-0.5cm}
    \includegraphics[width=\columnwidth,clip]{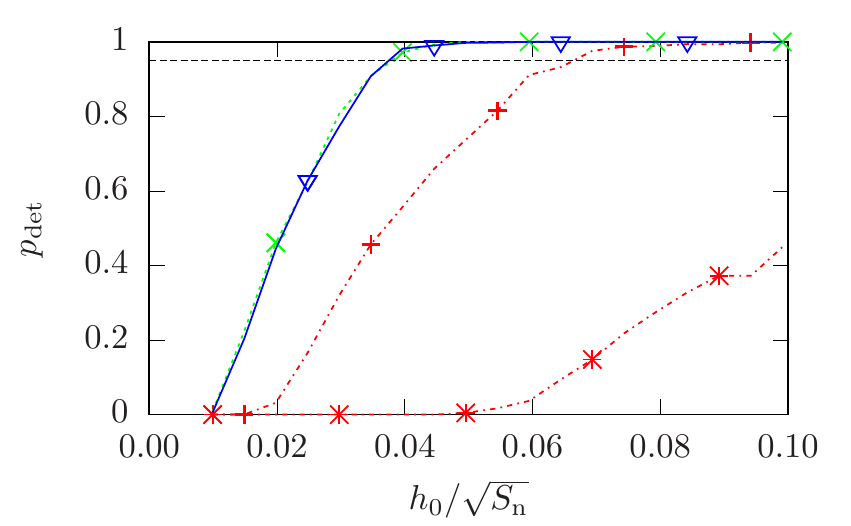}
    \raggedright (\dsc)\\\vspace*{-0.5cm}
    \includegraphics[width=\columnwidth,clip]{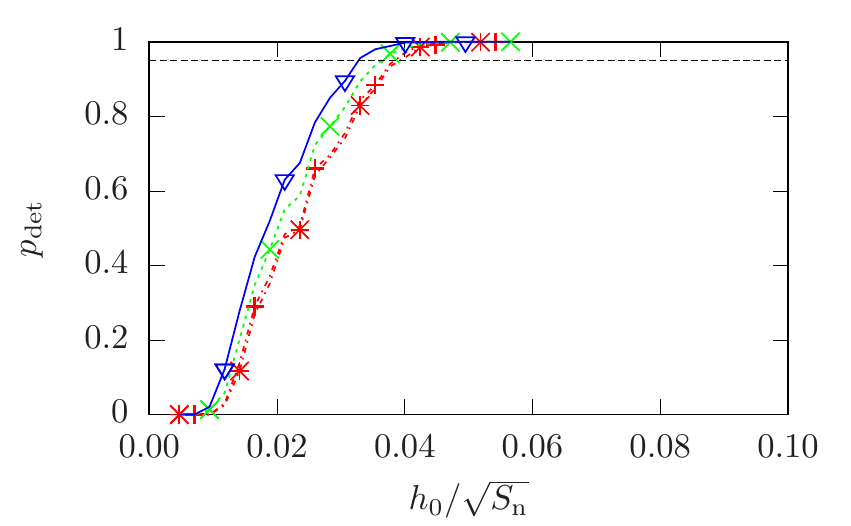}
  \caption{
    \label{fig:tests_realdata_detprobs_semicoherent}
    Detection efficiency $\pDet$ as a function of scaled signal amplitude $h_0/\sqrt{\Sn}$ for four
    different semicoherent statistics:
    $\scF$, 
    $\scFveto$, 
    $\utOSLsc$, 
    and $\OSNscpFA{-6}$. 
    Statistical errors are similar to the size of the symbols.
    The dashed horizontal line marks the $95\%$ detection probability level.
    The panels show:
    (\asc) a quiet band,
    (\bsc), (\csc) two bands with lines,
    (\dsc) a band with multiple disturbances. See Fig.~\ref{fig:tests_realdata_normSFT_semicoh} and
    Table~\ref{tbl:PsftthrSC} for more details on these data sets.
  }
\end{figure}


\subsubsection{Results for coherent statistics}
\label{sec:results-using-coher}

Figure~\ref{fig:tests_realdata_detprobs_coherent} shows the detection efficiency $\pDet$ as a function of the
scaled signal amplitude $h_0/\sqrt{\Sn}$, for the single-segment coherent statistics.

In the quiet band, shown in Fig.~\ref{fig:tests_realdata_detprobs_coherent}~(\acoh),
we find that the line-veto statistic $\utOSL$ has less detection power than the
$\F$-statistic, as would be expected since it does not match the noise population.
The conventional \FvetoName{} is safer than $\utOSL$ and performs just as well as the pure
$\F$-statistic.
The line-robust statistic $\OSN$ performs equally well as $\F$ and $\Fveto$ on this line-free data set.

In the disturbed bands shown in
Fig.~\ref{fig:tests_realdata_detprobs_coherent}~(\bcoh)-(\dcoh),
all statistics lose detection power to varying degrees.
We find that the \FvetoName{} is often able to recover most of the losses of the pure $\F$-statistic.
The line-veto statistic $\utOSL$ performs similarly in case (\bcoh) and yields an improvement over $\Fveto$ in
cases (\ccoh) and (\dcoh).
However, these cases show $\OSN$ to be more robust than either of the simpler vetoes.

Summarizing these results, we see that the line-robust statistic $\OSN$ consistently shows the best
performance over the different types of data: it is more robust to varying kinds of disturbances than $\Fveto$
and safer in Gaussian noise than $\utOSL$.

\subsubsection{Results for semicoherent statistics}
\label{sec:results-using-semi}

Figure~\ref{fig:tests_realdata_detprobs_semicoherent} shows the detection efficiency $\pDet$ as a function of
$h_0/\sqrt{\Sn}$ for the semicoherent statistics over the full data set.
Qualitatively, we find very similar results to the coherent case of
Fig.~\ref{fig:tests_realdata_detprobs_coherent}.

For the quiet band, shown in Fig.~\ref{fig:tests_realdata_detprobs_semicoherent}~(\asc),
we find that the simple line-veto $\utOSLsc$ loses a significant fraction of detection power compared to the
semicoherent $\scF$-statistic and to $\scFveto$, while the line-robust statistic $\OSNsc$ does not show any
significant degradation.

In the bands with noise disturbances
(Fig.~\ref{fig:tests_realdata_detprobs_coherent}~(\bsc)-(\dsc)),
it is again the $\scF$-statistic which suffers the most.
These examples show the line-robust statistic $\OSNsc$ consistently performing better than $\scF$ and as well
as or better than either $\utOSLsc$ or $\scFveto$ in all the disturbed bands.
The largest improvement is found in the example shown in
Fig.~\ref{fig:tests_realdata_detprobs_semicoherent}~(\csc), where the signal amplitude at $95\,\%$ detection
probability is nearly two times smaller for $\OSNsc$ compared to $\scFveto$.

\section{Conclusions}
\label{sec:conclusions}

We have extended the standard derivation of the $\F$-statistic by adding an explicit simple line hypothesis to
the standard Gaussian-noise hypothesis, namely a CW-signal-like disturbance in a single detector.
More work would be required to deal with coincident disturbances in multiple detectors.

Using the Bayesian framework we have derived two new detection statistics:
a ``line-veto'' statistic $\OSL$, which complements the $\F$-statistic and may be appropriate
for the follow-up of strong outliers, and a new line-robust detection statistic $\OSN$, which
contains both $\F$ and $\OSL$ as limiting cases.
We have also generalized both statistics to semicoherent searches.

The line-robust $\OSN$ requires choosing several prior parameters.
We have found in particular that the performance of $\OSN$ is sensitive to $\Ftho$, which regulates the
transition scale between $\F$ and $\OSL$.
This parameter stems from a rather unphysical prior in the $\F$-statistic derivation \cite{prix09:_bstat}, and
we could therefore only provide an ad-hoc empirical prescription for choosing it.
Further work to improve on this prior could also result in increased robustness when the detectors are not
equally sensitive.

The remaining parameters are more straightforward to interpret, as they encode the prior probability of line
artifacts.
For these we have tested both an ignorance prior and a simple adaptive tuning method.

We have tested the detection power of the new statistics on synthetic candidates, where both signal and noise
match our hypotheses, and on simulated signals injected into LIGO S5 data.
In both cases we have found that, with a reasonable choice of transition scale, $\OSN$ is consistently the
most robust in the presence of various types of instrumental artifacts.
In particular, it consistently equals or surpasses the performance of the popular ad-hoc
$\F$-statistic consistency veto, reaching up to a factor of two improvement in detectable signal
strength at 95\% confidence in example \csc{} in Fig.~\ref{fig:tests_realdata_detprobs_semicoherent}.

Combined with its close-to-optimal performance in undisturbed data, this makes $\OSN$ a promising statistic
for analyzing broadband, diverse data sets.

\section*{Acknowledgments}
This work has benefited from numerous discussions and comments from colleagues, in particular John T. Whelan,
Karl Wette, Evan Goetz, Berit Behnke, Heinz-Bernd Eggenstein and Thomas Dent.
We acknowledge the LIGO Scientific Collaboration for providing the data from the LIGO S5 run.
The injection studies were carried out on the ATLAS cluster at AEI Hannover.
P.L. and M.A.P. acknowledge support of the ``Sonderforschungsbereich'' Collaborative Research
Centre (SFB/TR7). D.K. was supported by the IMPRS on Gravitational Wave Astronomy.
This paper has been assigned LIGO document number \dcc{} and AEI-preprint number \aei{}.

\appendix

\section{Expectation value of the \texorpdfstring{$\F$}{F}-statistic under the line hypothesis \texorpdfstring{$\HypL$}{HL}}
\label{sec:expect-f-stat}

In order to derive the expectation value of the $\F$-statistic under the hypothesis $\HypL$ given in
Eq.~\eqref{eq:hypL}, we consider the more general case of a signal with detector-dependent amplitude
parameters $\Amp^\mu_X$.
The signal case corresponds to $\Amp^\mu_X = \Amp^\mu$ for all $X$, while the line hypothesis corresponds to
$\Amp^\mu_X = \Amp^\mu_Y \delta_{XY}$, with $Y$ denoting the detector containing the instrumental line.
By using the factorization of Eq.~\eqref{eq:Amuhmu}, we can express the data in the general case as
\begin{equation}
  \label{eq:6}
  x^X = n^X + \Amp_X^\alpha\,h^X_\alpha\,,
\end{equation}
with \emph{no} automatic summation over repeated detector indices $X,X',...$.
Hence, the projections of Eq.~\eqref{eq:xmuMmunu} for detector $X$ read as
\begin{equation}
  \label{eq:8}
  x^X_\mu \equiv \scalar{x^X}{h_\mu^X} = n^X_\mu + \Amp_X^\alpha \M^X_{\alpha\mu}\,,
\end{equation}
where we define $n^X_\mu \equiv \scalar{n^X}{h^X_\mu}$, and from Eq.~\eqref{eq:scalarproduct} we have
\begin{equation}
  \label{eq:9}
  x_\mu = \sum_X x^X_\mu\,,\quad\text{and}\quad
  \M_{\mu\nu} = \sum_X \M^X_{\mu\nu}\,.
\end{equation}
The expectation value of the $\F$-statistic of Eq.~\eqref{eq:Fstat} can therefore be written as
\begin{equation}
  \label{eq:11}
  \expect{2\F} = \M^{\mu\nu}\,\expect{x_\mu\,x_\nu}\,,
\end{equation}
where
\begin{align}
  \label{eq:12}
  \expect{x_\mu x_\nu} &= \sum_{XX'} \expect{x^X_\mu\,x^{X'}_\nu} \notag\\
  &= \M_{\mu\nu} + \sum_{XX'} \Amp_X^\alpha\M^X_{\alpha\mu} \,\M^{X'}_{\nu\beta}\Amp_{X'}^\beta\,,
\end{align}
using the noise expectation values $\expect{n^X_\mu n^{X'}_\nu} = \M^X_{\mu\nu}\,\delta^{XX'}$ for
uncorrelated noise between detectors, and $\expect{n_\mu^X} = 0$, assuming zero-mean noise $n^X$.

Given that $\M^{\mu\nu}\M_{\mu\nu}=4$, we obtain the general result for the expectation value of the
$\F$-statistic:
\begin{equation}
  \label{eq:13}
  \expect{2\F} = 4 + \M^{\mu\nu}\,\sum_{XX'} \Amp_X^\alpha\M^X_{\alpha\mu}
\,\M^{X'}_{\nu\beta}\Amp_{X'}^\beta\,.
\end{equation}
A CW signal has consistent amplitudes in all detectors, i.e., $\Amp_X^\alpha = \Amp^\alpha$,
and this leads to the well-known result for $\expect{2\F}$ in terms of the signal SNR $\snrS$:
\begin{equation}
  \label{eq:14}
  \expect{2\F}_{\HypS} = 4 + \snrS^2\,,\quad\text{with}\quad
  \snrS^2 = \Amp^\mu\M_{\mu\nu}\Amp^\nu\,.
\end{equation}
Under the line hypothesis $\HypL$, i.e., $\Amp_X^\mu = \Amp_Y^\mu\,\delta_{XY}$, we find
\begin{equation}
  \label{eq:15}
  \expect{2\F}_{\HypL} = 4 + \M^{\mu\nu}\,\Amp_Y^\alpha\M^Y_{\alpha\mu}\M^Y_{\nu\beta}\Amp_Y^\beta\,.
\end{equation}
In the special case of all detectors having identical antenna-pattern
matrices, i.e., $\M^X_{\mu\nu} = \M^Y_{\mu\nu}$, we have $\M_{\mu\nu} = \Ndet\,\M^Y_{\mu\nu}$, and
therefore
$\M^{\mu\nu} = \frac{1}{\Ndet}\M_Y^{\mu\nu}$. This results in
\begin{equation}
  \label{eq:16}
  \expect{2\F}_{\HypL} = 4 + \frac{1}{\Ndet}\,\snrL^2\,,
\end{equation}
with
\begin{equation}
  \label{eq:snrL}
  \snrL^2 \equiv \Amp_Y^\mu\M^Y_{\mu\nu}\Amp_Y^\nu\,,
\end{equation}
defining the (single-IFO) ``line SNR'' in detector $Y$.
The scaling of Eq.~\eqref{eq:16} should still approximately hold for detectors with not-too-different
antenna-pattern matrices $\M^X_{\mu\nu}$.

This result shows that a CW-like disturbance with SNR $\snrL$ in a single detector is not completely
suppressed in the multidetector $\F$-statistic, but is only reduced to an effective multidetector SNR of
approximately $\snrL/\sqrt{\Ndet}$.

\bibliography{../../biblio}

\end{document}